\documentclass[twocolumn, mathlines]{aastex631}

\usepackage{amsmath}

\usepackage{xcolor}

\begin{document}

\title{A pilot method to determine the high mass end of the Stellar Initial Mass Function in galaxies using UVIT, H$\alpha$-MUSE observations and applied to NGC628}

\author[0009-0005-6072-9252]{S Amrutha}
\affiliation{Indian Institute of Astrophysics, Koramangala II Block, Bangalore 560034, India};\correspondingauthor{Amrutha S}\email{amrutha.s@iiap.res.in}
\affiliation{Pondicherry University, R.V. Nagar, Kalapet, 605014, Puducherry, India}

\author[0000-0001-8996-6474]{Mousumi Das}
\affiliation{Indian Institute of Astrophysics, Koramangala II Block, Bangalore 560034, India}

\begin{abstract}

We present a pilot method to estimate the high-mass initial mass function (IMF) across the arm, interarm, and spur regions in galaxies and apply it to NGC 628. We extracted star-forming complexes (SFCs) from H$\alpha$ VLT/MUSE and UVIT (FUV and NUV) observations of NGC 628 and used ALMA observations to define the molecular gas distribution. We find that the extinction-corrected H$\alpha$ and FUV luminosities correlate well. Using the fact that O stars have a shorter lifetime (10$^7$ yr) compared to B stars (10$^8$ yr), we estimated the approximate number of O stars from H$\alpha$ emission, and the number of B0 ($M_{*} > 10 M_{\odot}$), and B1 ($10 M_{\odot} \geq M_{*} \geq 3 M_{\odot}$) stars using FUV, NUV observations. We derived the IMF index ($\alpha$) for different regions using O to B0 ($\alpha_{1}$) and B0 to B1 ($\alpha_{2}$) stellar ratios. Our findings indicate that if we assume H$\alpha$ arises only from O8-type stars, the resulting $\alpha_{1}$ value is consistent with the canonical IMF index. It steepens when we assume O stars with masses up to 100 $M_{\odot}$ with mean $\alpha_{1}= 3.16 \pm 0.62$. However, the $\alpha_{2}$ does not change for large variations in the O-star population, and the mean $\alpha= 2.64 \pm 0.14$. When we include only blue SFCs ($ FUV-NUV\leq0.3$), mean $\alpha_{2}$ is $2.43 \pm 0.06$. The IMF variation for SFCs in arms and spurs is insignificant. We also find that $\alpha_{2}$ correlates with different properties of the SFCs, the most prominent being the extinction-corrected UV color (FUV-NUV).

\end{abstract}


\section{Introduction} \label{sec:intro}

Stars in galaxies are generally found in star clusters and stellar associations \citep{1986FCPh...11....1S}. The masses of the stars formed in a cluster vary over a wide range and depend on several factors, such as the environment, metallicity, and density of the parent molecular cloud. This distribution of stellar masses formed during the star-formation event in a given volume is called the initial mass function (IMF) \citep{2002Sci...295...82K}. The IMF is a crucial ingredient for the stellar evolution models of star-forming galaxies (e.g., \cite{1999ApJS..123....3L, 2003MNRAS.344.1000B, 2005MNRAS.362..799M}) and is thus essential for understanding star formation as well as galaxy formation and evolution \citep{Bastian_2010}. 

Over the past few decades, many studies have tried to understand the IMF using resolved populations of stellar clusters in the solar neighborhood, within the Milky Way, and in nearby galaxies lying in the Local Group \citep{1955ApJ...121..161S, 2015ApJ...806..198W, Wainer_2024}. The IMF is commonly described by a power-law form with an index $\alpha$ =2.35, where dN/dm= Am$^{-\alpha}$, A is a constant, and dN/dm represents the number of stars formed in a mass range dm  \citep{1955ApJ...121..161S}. Even the canonical stellar IMF is closer to value 2.3 for stars with $M_{*}> 0.5 M_{\odot}$. This IMF is observed for the field stars and nearby star-forming regions of sizes approximately equal to one parsec \citep{2018A&A...620A..39J}. In general $\alpha$ is observed to be shallower for low mass stars ($M_{*}< 0.5 M_{\odot}$) \citep{2001MNRAS.322..231K, Chabrier_2003}. Determining the uniformity of the IMF for all stellar populations involves combining IMF estimates for different resolved populations. However, the estimation of IMF in different mass ranges has its own challenges \citep{2002Sci...295...82K}.

Several studies have tried to understand the universality of the IMF \citep[e.g.,][]{2008ApJ...675..163H, Meurer_2009, 10.1093/mnras/stu1521}. Recent studies have shown that the IMF is non-universal and depends on environmental and statistical effects \citep{ 2008Natur.455..641P, KRUMHOLZ201449, 2022MNRAS.509.1959S, 2022MNRAS.516.5712T}. The denser starburst galaxies have a top-heavy IMF, i.e., a large number of massive stars and hence an IMF that is flatter than the canonical IMF. In comparison, the less dense, low surface-brightness (LSB) galaxies show a bottom-heavy IMF (i.e., steeper than the canonical IMF) \citep{10.1007/978-94-009-0543-6_18, 10.1111/j.1365-2966.2010.17959.x, 2008ApJ...675..163H, Meurer_2009}. Even within galaxies, there are regions with low stellar surface mass densities, such as the outer disks of extended UV (XUV) galaxies or the inter-arm regions of spiral galaxies where the IMF can be different compared to that of the spiral arms \citep{2007ApJS..173..538T, kroupa2024initialmassfunctionstars}.

In this study, we focus on the high-mass end of the IMF. Massive stars significantly impact the chemical enrichment of the interstellar medium (ISM) due to their strong stellar winds and feedback process \citep{Chappell:1999fr, Freyer_2003}. Hence, it is essential to understand the upper or high-mass end of the IMF, even though the low-mass stars contribute more to cluster masses as well as the overall mass of a galaxy. Young massive stars emit most of their energy at far-ultraviolet (FUV) wavelengths and have a short main sequence. However, it is difficult to distinguish between the two main types of massive stars (O and B, $M_{*}> 20 M_{\odot}$) based on just magnitude alone \citep{2002Sci...295...82K}. Even with UV colors, it is challenging to distinguish massive B stars from O stars \citep{koda.etal.2012}. Massive O-type stars give out FUV, NUV, and H$\alpha$ emission. The H$\alpha$ emission from O and massive B stars is a well-known tracer of star formation in galaxies \citep{2009ApJ...703.1672K, Lee_2009}. But H$\alpha$ can trace star formation for only 10 Myr, as O star lifetimes vary from 1 to 10 Myr. However, UV traces O, B, and even some red and cold stars, which makes them trace star formation for around 200 Myr \citep{2005ApJ...619L..79T}. Therefore, recent studies have used both UV and H$\alpha$ measurements to understand the universality of the high mass end of the IMF by considering the relative strength of their emission in external galaxies \citep{Meurer_2009, Lee_2009, koda.etal.2012}. 

Previous studies have used Galaxy Evolution Explorer (GALEX) data to show that there is a correlation between the star formation rate (SFR) and the H$\alpha$ to UV flux ratios in galaxies \citep{Meurer_2009, Lee_2009, Hunter_2010}. They also show that the galaxies with low SFRs have a deficit of O stars and hence have bottom-heavy IMF due to stochastic sampling. \cite{Boselli_2009} used H$\alpha$ to UV ratios to show that the high mass end of late-type normal galaxies follows the standard IMF. \cite{2010ApJ...719L.158C} showed that a universal IMF exists by comparing the H$\alpha$ flux with the cluster mass. Some studies have investigated the high end of the IMF in the outskirts of XUV galaxies and find that the bottom-heavy IMF dominates \citep{Gogarten_2009, 10.1093/mnras/stz3151}. In contrast, some studies show that it follows the standard IMF \citep{koda.etal.2012}. A few studies analyzed the IMF in XUV disks by assuming that the H$\alpha$ to UV flux ratios are approximately equal to the ratios between the lifetime of the stars that can produce H$\alpha$ to those that produce UV emission (i.e., mainly O and B types stars) \citep{2007AJ....134..135Z}. \cite{10.1093/mnras/stz3151} have used HST measurements and compared them with the simulations to obtain the high mass end of the IMF.

This paper presents a new method to determine the slope $\alpha$ of the high mass end of the IMF using observations of individual star-forming complexes (SFCs). We leverage the high sensitivity and spatial resolution of the Ultraviolet imaging telescope (UVIT) and Multi Unit Spectroscopic Explorer (MUSE) to extract the SFCs of a nearby galaxy to determine the ratio of O and B stars to derive $\alpha$. We investigate whether $\alpha$ varies with regions, i.e., spiral arms, spurs, and inter-arm regions, as the mechanisms leading to star formation in these regions are different. Spiral arms are global disk instabilities, whereas star formation in the spurs or feathers is due to local gas instabilities, where diffuse gas is transformed into dense gas when it flows into the spiral arm. The inflow shocks the gas near the arm and causes star formation in these spur regions \citep{2002ApJ...570..132K, 2006ApJ...647..997S}. The inter-arm regions have isolated SFCs, and the star-forming mechanism may be more local. Many studies have attempted to understand star formation in both the spiral arm and inter-arm regions, but our analysis focuses on the IMF \citep{2001Ap&SS.276..517S, Foyle_2010, 10.1093/mnras/stt279}. Studying the IMF over different regions of a galaxy will also help us understand the universality of the IMF within a galaxy.

To test our new method, we applied it to the nearby face-on galaxy NGC 628. In Section \ref{NGC 628 and Observations Data}, we briefly introduce the NGC 628 and discuss the data. Section \ref{sec:Data Analysis} discusses the methods used to extract the complexes, correct for the extinction, and the characterization of the SFCs. In Section \ref{sec:Results} and \ref{sec:discussion}, we present and discuss the results of our IMF study over different regions and how it depends on the properties of the SFCs. 


\begin{figure*}[ht!]
\includegraphics[scale=0.14]{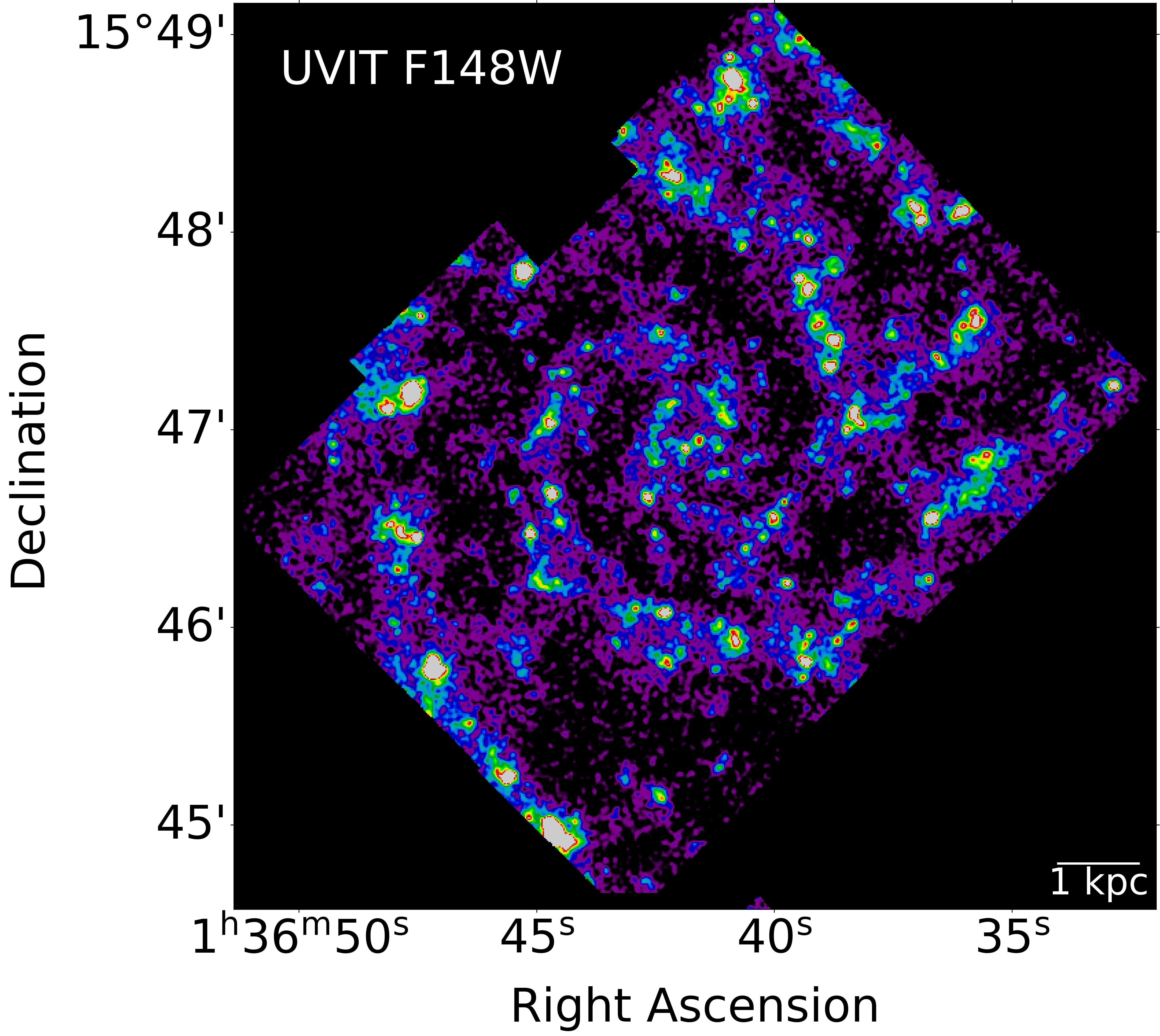}
\includegraphics[scale=0.14]{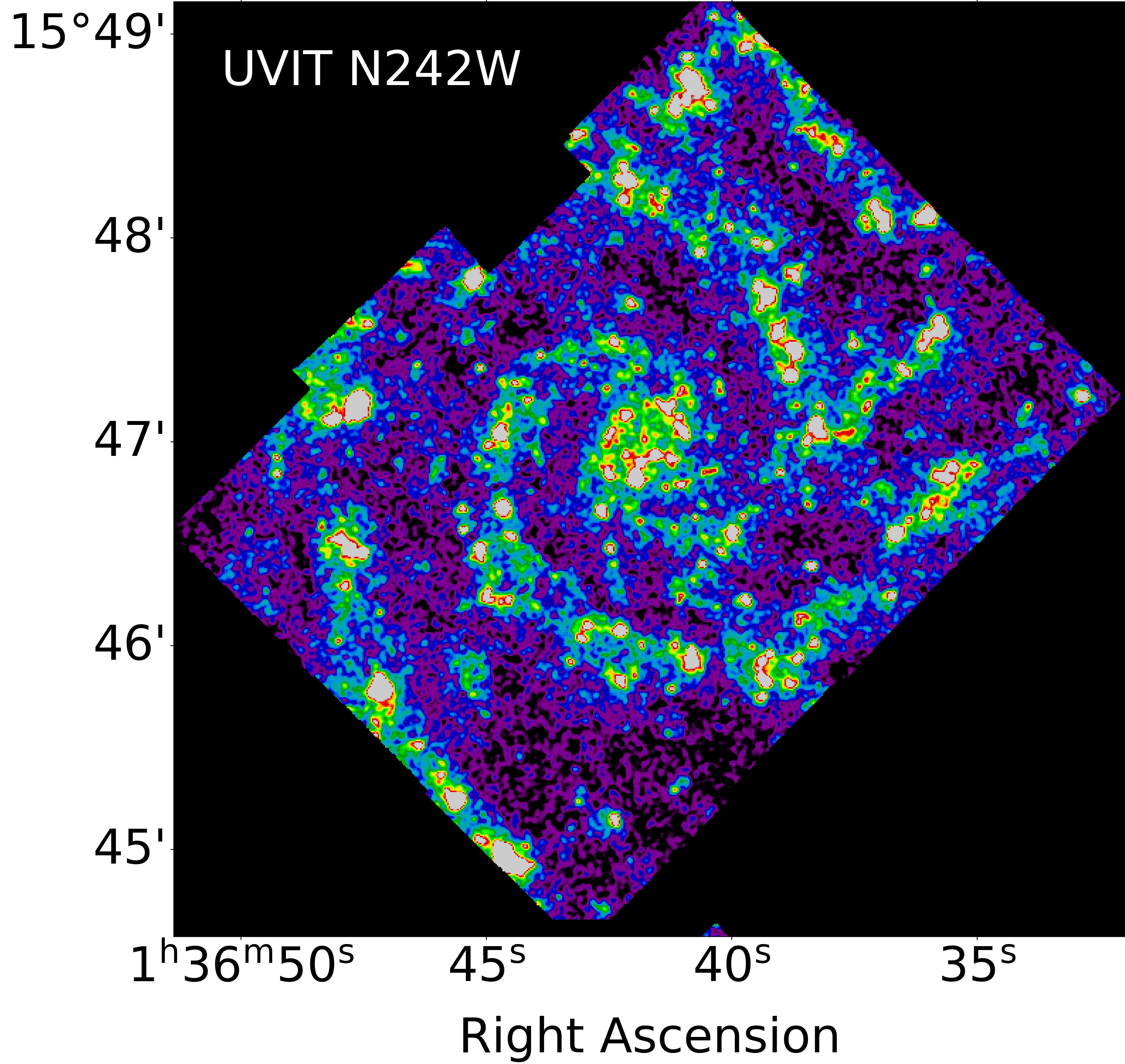}
\includegraphics[scale=0.14]{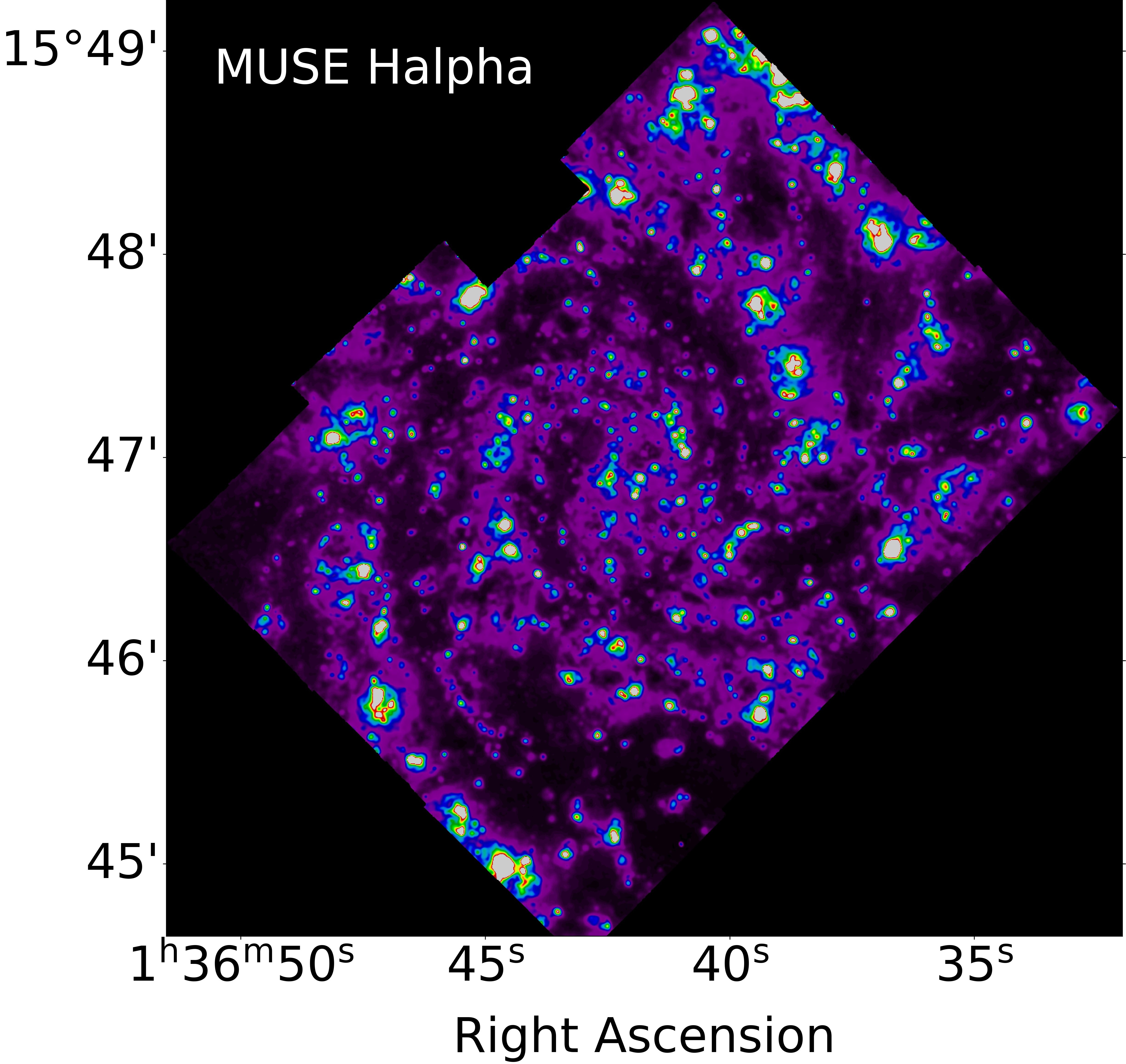}
\caption{UVIT FUV, NUV, and MUSE H$\alpha$ images of galaxy NGC 628 tracing recent star formation.}
\label{fig1}
\end{figure*}

\section{NGC 628 and Observations} \label{NGC 628 and Observations Data}
NGC 628, also known as M74 and the Phantom Galaxy, is a grand design spiral galaxy of Hubble type SA(s)c \citep{1991rc3..book.....D}, and is at a distance of 9.84 Mpc \citep{10.1093/mnras/staa3668}. This massive galaxy (log($M_{*}/M_{\odot}$)=10.34; \citep{2019ApJS..244...24L}) is nearly face-on (i=8.9\textdegree ;\cite{Lang_2020} ) with a position angle of 20.7\textdegree \citep{Lang_2020}. Hence, it is the perfect target for studying star formation properties in different regions of a galaxy. The star formation in NGC  628 is well studied \citep{ Kreckel_2016, Kreckel_2018, 10.1093/mnras/stac2940} and the SFCs on kpc scales have also been characterized \citep{Yadav_2021, Ujjwal_2022}. We used archival, deep UV imaging observations obtained using the UVIT on board the AstroSat telescope \citep{2012SPIE.8443E..1NK} to understand IMF. The UVIT is a twin telescope with co-aligned Ritchey-Chrétien (RC) optics. The FUV (1300–1800Å) is one of the telescopes, and the other has the near-UV (NUV; 2000–3000Å) and visible (VIS) bands. The instrument has a field of view of 0.5\textdegree and can observe simultaneously in all three bands. The FUV and NUV have a resolution around \(1.2"- 1.5"\) sampled at 0.417$^{\prime\prime}$ per pixel \citep{Tandon_2020}. This resolution is three to four times better than  GALEX \citep{2005ApJ...619L...1M}. 

We downloaded Level 1 data from the Indian Space Science Data Centre (ISSDC) website\footnote{\fontsize{6pt}{12pt}\selectfont \url{https://astrobrowse.issdc.gov.in/astro_archive/archive/Home.jsp}}. The UVIT observed NGC 628 in multiple filters, but we used the CaF2 F148W FUV image with an exposure time of 1810s and a bandwidth of 500Å, and the NUVB4 N263M image, which has an exposure time of 2086s and a bandwidth of 275Å, for our analysis. We also used two close filters, F172M (bandwidth and exposure time are 125Å and 5881s respectively) and N219M (bandwidth and exposure time are 270Å and 1365s respectively), to correct internal extinction using the Beta slope as described in Section \ref{Internal extinction correction}.

We obtained archival Multi Unit Spectroscopic Explorer (MUSE; \cite{Eric}) H$\alpha$ and H$\beta$ maps, archival Atacama Large Millimeter/submillimeter Array (ALMA; \cite{Leroy_2021a, Leroy_2021b}) CO(J=2 - 1) moment 0 map, and also the James Webb Space Telescope (JWST; \cite{Lee_2023}) F2100W map, from the PHANGS Treasury Program\footnote{\fontsize{6pt}{12pt}\selectfont \url{https://sites.google.com/view/phangs/home}}. The Very Large Telescope (VLT) MUSE observations provide a field of view of one arcmin $\times$ one arcmin, with a high spatial resolution of 0.6$^{\prime\prime}$ PSF FWHM sampled at 0.2$^{\prime\prime}$ per pixel. Whereas ALMA has a resolution around 1$^{\prime\prime}$ sampled at 0.2$^{\prime\prime}$ per pixel. The JWST F2100W map has a central wavelength approximately equal to 21$\mu m$ and a resolution of 0.67$^{\prime\prime}$ sampled at 0.11$^{\prime\prime}$ per pixel.

\section{Data Analysis} \label{sec:Data Analysis}

\subsection{Source extraction and classification} \label{Source extraction and classification}

We reduced the raw UVIT data using the graphical user interface, CCDLAB \citep{2017PASP..129k5002P}. This interface corrects field distortions, flat-fielding, and drift. Later, it combines the orbit-wise UVIT images to create a final deep image. We then did the astrometry using the same interface. We did background subtraction of these images using IRAF as mentioned in \cite{10.1093/mnras/stae907}. Using around 5 to 6 stars in the FUV and NUV image fields, we found that the mean PSF of the field stars is around 1.2$^{\prime\prime}$. Hence, this is assumed to be the resolution of the UV images. This resolution corresponds to a physical scale of 57 pc at the distance of NGC 628, which is moderately high resolution and cannot resolve star clusters as clusters have radii of a few parsec \citep{2015MNRAS.452..525R}. However, we can resolve OB associations and SFCs with this resolution, as they correspond to a scale of a few hundred parsec \citep{Elmegreen_2014, 10.1093/mnras/stae907}.

Since we are interested in comparing the H$\alpha$ and UV emission, we made cutouts of the UVIT FUV and NUV images that match the size of the MUSE H$\alpha$ map. The cutout images of the FUV, NUV, and H$\alpha$ maps are shown in Figure \ref{fig1}. We convolved the H$\alpha$ map to the UV resolution of 1.2$^{\prime\prime}$. We extracted SFCs in all three images using the command line program Source Extractor (SExtractor; \cite{1996A&AS..117..393B}). We set the detection threshold to five times the estimated global background noise for confident detection. 
Then, we masked the foreground sources (with high parallax and high proper motion) by comparing the extracted SFCs with the bright sources in the Gaia catalog using TOPCAT \citep{2005ASPC..347...29T}. We also convolved the H$\beta$ and JWST map to the UVIT resolution.

We determined the total counts from the extracted SFCs using the Photutils module \citep{2018AJ....156..123A}. The UV fluxes were derived using the conversion relation  \(Flux=UC \times CPS_{corr}\) as mentioned in \cite{2017JApA...38...28T}, where UC is unit conversion factor in \(ergs^{-1}cm^{-2}A^{-1}\) and  \(CPS_{corr}\) is the counts per second corrected for Milky Way extinction \citep{2011ApJ...737..103S} for the FUV and NUV filters \citep{2008A&A...492..277B}. We determined the UV and H$\alpha$ luminosities in \(ergs^{-1}\) units using the distance of the galaxy and the bandwidth of the UV filters. 

We matched the UV and H$\alpha$ images and found the SFCs that are present in all three filters. We also determined the SFCs that are detected only in the FUV and H$\alpha$ filters, the SFCs detected only in the NUV and H$\alpha$ filters, and finally, the SFCs detected in only the FUV and NUV filters. We also found the SFCs that are visible only in FUV, only in NUV, and also only in H$\alpha$. This classification was necessary as we needed emission from all three bands to estimate the IMF value.

\begin{figure*}[ht!]

\includegraphics[scale=0.63]{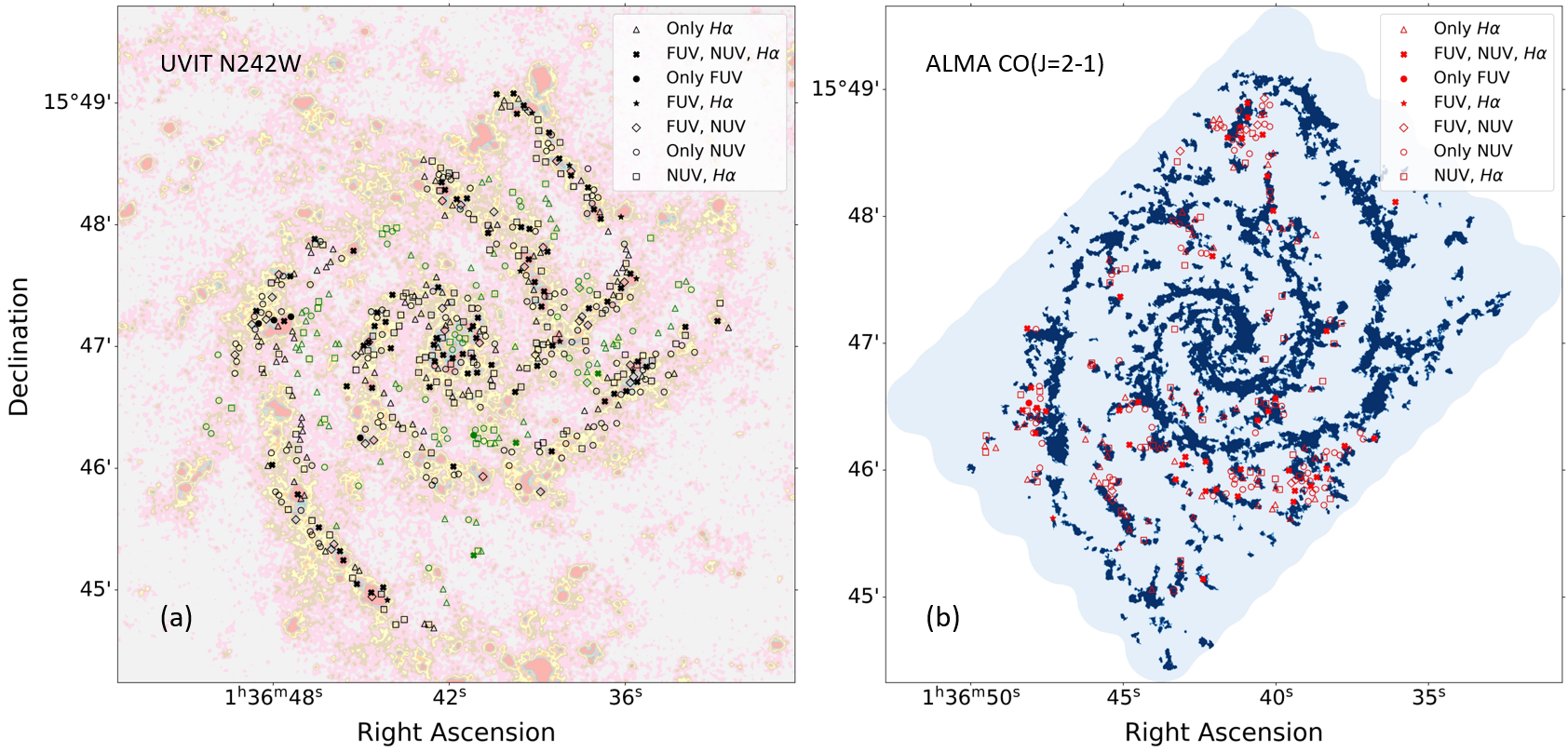}
\caption{The above images show the classification of SFCs based on their region(different colors) and their emission in different filters (different markers). Image (a):  The locations arm (black markers) and interarm (green markers) SFCs overplotted on UVIT NUV image. Image (b): The locations of spurs(red markers) SFCs overplotted on the ALMA CO(J=2-1) image.}
\label{fig_regions}
\end{figure*}

We noted the SExtractor parameters of the SFCs with larger areas present in at least two filters. We also classified them as arm, interarm, and spur SFCs based on location. This classification helps us understand the universality of the IMF within the galaxy. Hence, we first classified the SFCs as arms and interarm sources based on their location in the NUV image (see Figure \ref{fig_regions}a). Then, we again checked which SFCs lie in and around the spur regions of the ALMA observations (Figure \ref{fig_regions}b).  

\subsection{Internal extinction correction} \label{Internal extinction correction}

\begin{figure*}[ht!]
\includegraphics[width=1\linewidth]{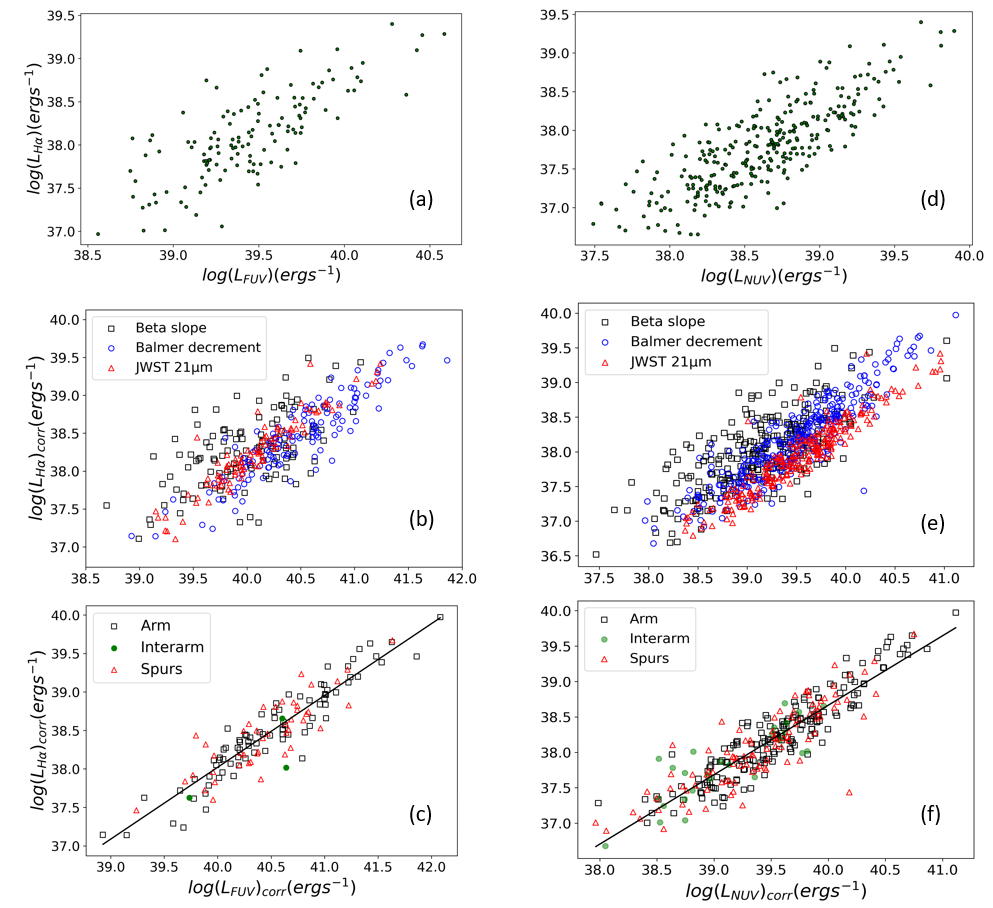}
\caption{Figures on the left (a, b, and c) and on the right (d, e, and f) illustrate the correlation of H$\alpha$ luminosity with FUV and NUV luminosities, respectively, for the SFCs. Figures a and d: Luminosities presented without any correction for internal dust attenuation. Figures b and e: Luminosities are corrected for internal dust attenuation using different methods indicated with different symbols and colors. Figures c and f: Luminosities corrected for internal dust attenuation using the Balmer decrement method for the SFCs with H$\alpha$ emission, and the scaling relation is used to find the FUV extinction coefficient. The fitted relations related to these plots are given in Section \ref{FUV and Halpha luminosity correlation}.}
\label{fig2}
\end{figure*}

An accurate extinction correction of observed H$\alpha$ and UV luminosities is crucial for IMF analysis. As there are different methods to correct the galaxy's internal extinction, we corrected SFC luminosities for internal extinction using the Balmer decrement, Beta slope, and JWST 21$\mu m$ methods to see which method gives a better correlation for H$\alpha$ and UV luminosities. We used H$\alpha$ and H$\beta$ maps of MUSE to find the Balmer decrement. It is one of the most successful techniques for determining dust extinction. It uses the ratio of two nebular Balmer emission lines like H$\alpha$ and H$\beta$ at low redshifts \citep{1992ApJ...388..310K, Moustakas_2006, 10.1111/j.1365-2966.2010.17321.x}. The intrinsic ratio of the two lines remains roughly constant for typical gas conditions in star-forming galaxies and $(\frac{H\alpha}{H\beta})_{int}=2.86$, at the temperature $10^{4}$ K and an electron density $n_{e}=10^{2}cm^{-3}$ for Case B recombination \citep{1989agna.book.....O}. Hence, the corresponding color excess $E(B-V)$ is given by,
\begin{equation} \label{eq1}
 E(B-V)=1.97log_{10}[\frac{(\frac{H\alpha}{H\beta})_{obs}}{2.86}]
\end{equation}

\noindent
and using the \cite{2000ApJ...533..682C} reddening curve, we find the $H\alpha$ extinction to be, $A_{H\alpha}=(3.33 \pm 0.80) \times E(B-V)$.

For the SFCs with FUV and H$\alpha$ emission, we found FUV extinction using the relation, $A_{FUV}=3.6A_{H\alpha}$, assuming that the FUV emission arises mainly from the young stellar population as mentioned in \cite{2008AJ....136.2782L}, and we corrected for the extinction in NUV band using the relation $A_{NUV}=0.8A_{FUV}$ \citep{2000ApJ...533..682C}. The extinction-corrected H$\alpha$ and UV luminosity is given by
\begin{equation} \label{eq:corr_luminosity}
L({\lambda})_{corr}=10^{0.4A_{\lambda}}L({\lambda})_{obs}
\end{equation}

where $L({\lambda})_{obs}$ is observed luminosity at the wavelength $\lambda$. The correlation of extinction corrected H$\alpha$ and UV luminosity for SFCs using the Balmer decrement method is shown in Figure \ref{fig2}c and \ref{fig2}f.

The Beta slope method can be used for the SFCs with FUV and NUV emission to correct the internal extinction \citep{1994ApJ...429..582C}. The extinction curve in the UV band (1300-2600 Å) can be fitted by a power law with the slope $\beta$ ($f_{\lambda} \propto \lambda^{\beta}$). The UV spectral slope is filter-dependent, and we can get a better estimate for $\beta$ when we consider the two close filters. Hence, $\beta$ for UVIT filters $m_{F172M}$ and $m_{N219M}$ is given by,
\begin{equation} \label{eq2}
 \beta=\frac{m_{F172M}-m_{N219M}}{-2.5log(\frac{\lambda_{F172M}}{\lambda_{N219M}})}-2
\end{equation}
where $m_{F172M}$ and $m_{N219M}$ are the Galactic extinction corrected magnitudes of SFCs in the F172M and N219M filters with $\lambda_{F172M}$ and $\lambda_{N219M}$ as their central wavelengths respectively. Then, we used filters F148W and N263M and found the SFC's extinction $A_{\lambda}=(0.44\pm 0.03)E(B-V)k(\lambda)$ as given by \cite{2000ApJ...533..682C}. The color excess E(B-V) is then found using $\beta=-2.616+4.594E(B-V)$ as given by \cite{Reddy_2018} and $k(\lambda)$ as given in \cite{2000ApJ...533..682C} for filters F148W and N263M, and for SFCs with H$\alpha$ emission. We again calculated extinction corrected H$\alpha$ and UV luminosities for this method using equation \ref{eq:corr_luminosity}.

Apart from the two methods mentioned above, we use 24$\mu m$ to measure the embedded H$\alpha$ and UV luminosities, as mid-IR peaks around 24$\mu m$ and traces the hot dust component that is due to the young stellar population. Also, \cite{refId0} showed that the F2100W filter of JWST is better for extinction correction than any of the other mid-IR bands in JWST. Hence, here we corrected for extinction using a convolved JWST 21$\mu m$ map. We used the extinction correction relation for H$\alpha$ as given in \cite{2009ApJ...703.1672K} and for FUV and NUV as in \cite{Hao_2011}. The relation for finding the extinction-corrected luminosity is $L(\lambda)_{corr}=L(\lambda)_{obs}+cL(24 \mu m)$. Here $L(\lambda)_{corr}$ and $L(\lambda)_{obs}$ are the corrected luminosity and observed luminosity in $ergs^{-1}$ respectively. The parameter c is 3.89, 2.26, and 0.0024 when we correct the FUV, NUV, and H$\alpha$ luminosities, respectively. $L(24 \mu m)$ is the Spitzer MIPS 24$\mu m$ luminosity in $ergs^{-1}$. Since we are using the JWST 21$\mu m$ map, we took the ratio of JWST 21$\mu m$ and MIPS 24$\mu m$ luminosity for some of the SFCs on both maps. We find that the ratio is approximately equal to 4.54. We multiplied this value with $L(21 \mu m)$ to obtain the $L(24 \mu m)$. 

We plotted the extinction corrected  H$\alpha$ and UV luminosity correlation for all three methods in Figure \ref{fig2}b and \ref{fig2}e. The extinction-corrected luminosities strongly correlate for SFCs corrected with the Balmer decrement and JWST methods. However, we see the clear difference between the three methods in Figure \ref{fig2}e, which reveals that each method uniquely compensates for extinction at varying depths. Hence, it is necessary to use a single method to correct extinction. The SFC luminosities corrected with the Beta slope method show more scatter ($\sigma_{FUV}$=0.38 and $\sigma_{NUV}$=0.37), making it less preferable. Although the JWST 21$\mu m$ method showed a tight H$\alpha$ and UV luminosity correlation ($\sigma_{FUV}$=0.17 and $\sigma_{NUV}$=0.17), due to its limited field of view, very few SFCs had  21$\mu m$ emission. Hence, we mainly used the Balmer decrement method ($\sigma_{FUV}$=0.21 and $\sigma_{NUV}$=0.25) for the extinction correction for SFCs with H$\alpha$ luminosity; the SFCs without H$\alpha$ luminosity were not used for further analysis wherever H$\alpha$ and UV luminosities are involved. However, we have shown the extinction corrected H$\alpha$ and UV luminosity correlation for the Balmer decrement method and also the SFR correlations for the Balmer decrement and JWST methods in Section \ref{FUV and Halpha luminosity correlation}.

\subsection{Characterization of the SFCs} 
\label{Characterization of the SFCs}

\begin{figure*}[ht!]
\centering
\includegraphics[scale=1]{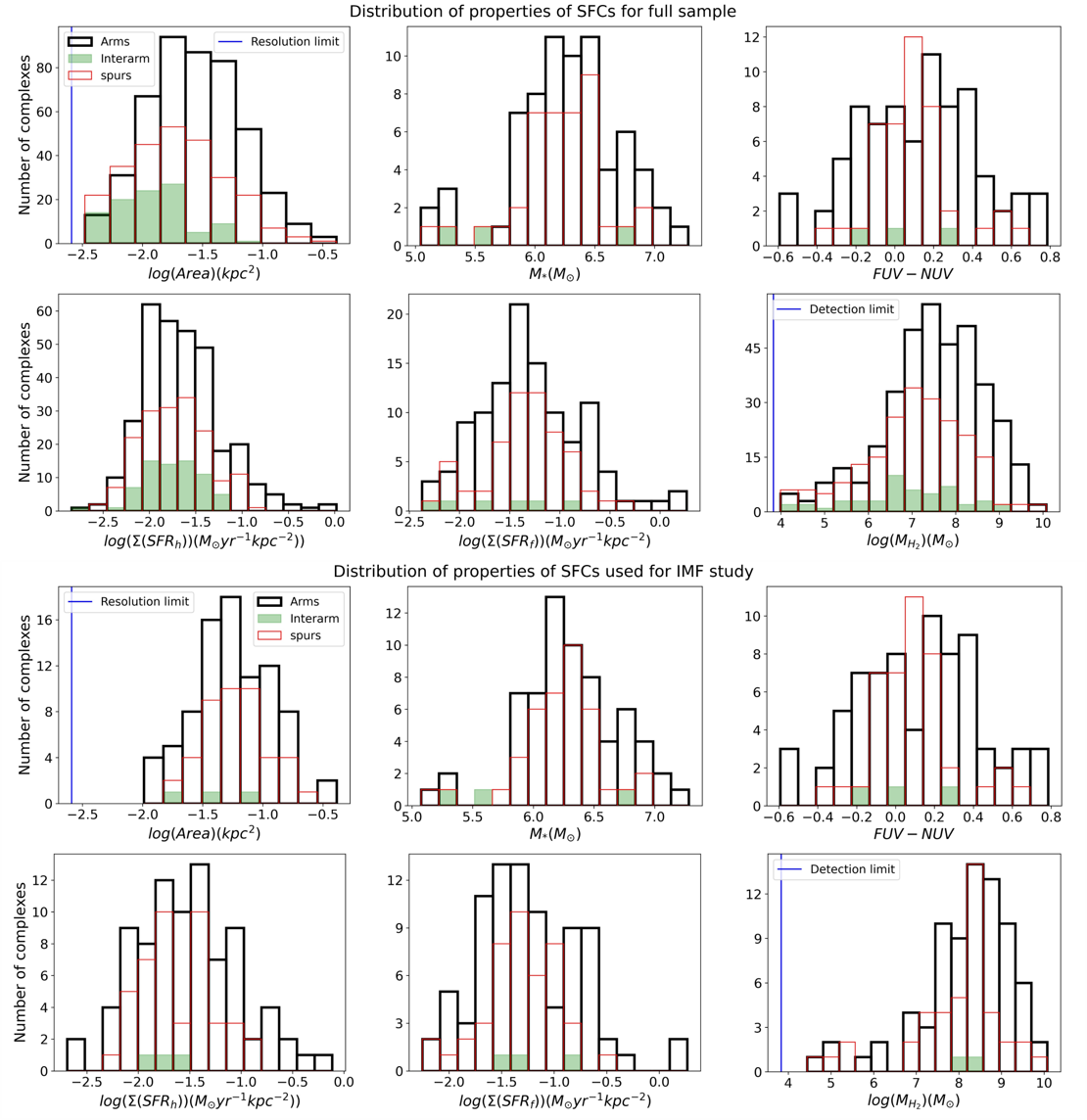}

\caption{The distribution of star formation properties across different regions (arms, spurs, and interarm) for all available SFCs (first two rows) and specifically for the SFCs used in the IMF analysis.}
\label{fig3}
\end{figure*}

We calculated the FUV SFR for the SFCs using \cite{Salim_2007}. The FUV SFR in \(M_\odot yr^{-1}\) is given by $SFR_{FUV}=0.68 \times 10^{-28} \times L(FUV)_{corr}$. Similarly, we calculated the H$\alpha$ SFR using \cite{Calzetti_2007} given by $SFR_{H\alpha}=5.3 \times 10^{-42} \times L(H\alpha)_{corr}$. Here \(L(FUV)_{corr}\) and \(L(H\alpha)_{corr}\) are the  extinction corrected luminosities in \(ergs^{-1}\). As mentioned in Section \ref{Internal extinction correction}, we only derived both FUV and H$\alpha$ SFRs for SFCs with associated H$\alpha$ luminosity. We calculated the SFR density \(\Sigma(SFR)\) by dividing the SFRs by their respective areas.

We also derived the corresponding stellar masses of the SFCs using starburst99 and a simple stellar population model (SSP) \citep{1999ApJS..123....3L}. The starburst99 is a set of spectrophotometric model predictions that can be used to find the properties of star-forming galaxies. We ran starburst99 assuming Padova tracks with asymptotic giant branch stars and standard Salpeter IMF \citep{1955ApJ...121..161S} with the lower cutoff mass set to $m_{l} = 0.1M_{\odot}$ and the upper cutoff mass to be $m_{u} = 100M_{\odot}$. We adopted a solar metallicity (Z=0.02; \cite{1999ApJS..123....3L}), which is closer to the value mentioned in \cite{2015AJ....149...51C}. We evolved the model from 1 Myrs to 200 Myrs, as UV can trace up to 200 Myrs. We plotted the observed extinction-corrected UV color and FUV magnitudes of the SFCs on the theoretically predicted values from Starburst99. We then interpolated the values for each SFC to find the mass.

We estimated the molecular hydrogen mass using the ALMA CO(J=2-1) map. We converted the CO flux to $H_{2}$ mass using the conversion factor $\alpha_{CO(}=4.35M_{\odot} pc^{-2}(Kkms^{-1})^{-1}$ and $R_{21}=0.62$ \citep{2013ApJ...777....5S, Kreckel_2018}.

We plotted the distribution of different properties of SFCs, like the area (from SExtractor), molecular hydrogen mass ($M_{H_2}$), extinction-corrected FUV and NUV color (FUV-NUV), stellar mass ($M_{*}$), and  \(\Sigma(SFR)\) for both H$\alpha$ and FUV emission for all the possible SFCs (For area and $M_{H_2}$ distribution, we included even the SFCs which did not have emission in all three bands) and also for the SFCs used for IMF analysis as shown in the Figure \ref{fig3}. We also observed that the median estimated values of properties like ($M_{*}$) and area of SFCs are more than what we see in resolved studies of stellar clusters in NGC 628 with HST in Legacy ExtraGalactic UV Survey with the Hubble Space Telescope (LEGUS; \cite{2017ApJ...841..131A}) as SFCs contains multiple individual star-forming clusters and smaller OB associations.

\subsection{The number of high-mass stars and their ratios in the SFCs} \label{Counts of High mass stars and their ratios}

This section describes how we estimated the approximate number of O and B stars in the SFCs. It depends crucially on the difference in time scales over which H$\alpha$ and UV emission trace star formation. The $H\alpha$ and FUV emission arises from O stars and massive B stars. But since H$\alpha$ comes from recombination in the ionization region around massive stars, it lasts for a few Myr. Even if we consider that there is a contribution from B stars in the H$\alpha$ emission, we may not get the total count of massive B stars. Hence, for this study, we assume that the $H\alpha$ emission arises mainly from O stars. The FUV emission arises from O (M $ > 21 M_{\odot}$), and also massive B0 stars (M $ > 10 M_{\odot}$). We also see that with a 5$\sigma$ photometric limit of FUV and NUV filters, we can only detect individual stars after O7 stars. But, since low-mass stars are more in number, and studies have shown that NUV can significantly detect emission from even low-mass B stars \citep{Das_2021}, we assume that the NUV emission arises from O, B0, and B1 stars (M $\leq 10 M_{\odot}$). All of this information can be used to estimate the number of O, B0, and B1 stars as described below. 

As we do not know what kind of O stars produce the $H\alpha$ emission from the SFCs, we randomly populate O stars in the SFCs by assigning a fraction of the $H\alpha$ luminosity for each type of O star we are considering. Here, we present the three cases. (i)~First, we consider that $H\alpha$ arises from only O8V-type stars (31 M$_{\odot}$). (ii)~Secondly, we consider that $H\alpha$ comes from only O7V, O8V and O9V stars. We included these types of O stars because previous studies show that the median $H\alpha$ luminosity of the SFCs in NGC 628 corresponds to the luminosity of a single O8V star \citep{Kreckel_2016}. (iii)~Apart from these two combinations, we tried a third case in which we populate the SFCs with a combination of O9V (23 M$_{\odot}$) to O3V stars (88 M$_{\odot}$). In this way, we include different combinations of O-type stars to estimate how the IMF varies. We used the $H\alpha$ Lyman photon flux $Q_{H}$ for each stellar type from \cite{Sternberg_2003}. We consider Case-B recombination with electron temperature $10^{4} K$ \citep{2002ApJ...576..135G}. The $H\alpha$ luminosity in $ergs^{-1}$ is given by $1.37 \times 10^{-12}Q_{H}[s^{-1}]$. The \cite{castelli2004newgridsatlas9model} model was used to find the FUV and NUV luminosities for these stars. We chose the closest parameters, such as effective temperature ($T_{eff}$), surface gravity (g), and radius (R) from the grid of all types of O, B0, and B1  star models from \cite{Sternberg_2003} and \cite{ 2010ApJS..189..309L}, where the B1 stars parameter are from the second paper and the rest from the first paper. The parameters ($T_{eff}$, log(g), R), along with the UV luminosities for the solar metallicity, are given in Table \ref{table1}. 


Initially, the algorithm carried out 10,000 iterations for each SFC, drawing a random $H\alpha$ luminosity probability for each type of O star from a uniform distribution between 0 and 1. For the third case, we ensured that there should be at least one of each type of O star in a SFC. Then, the total number of O stars present in a SFC for each of the three cases is,

\begin{equation} \label{eq3}
N(O)= \sum_{n=3}^{9} \frac{P(On) \times L_{H\alpha}}{L_{H\alpha}(On)}
\end{equation}

\noindent
where, $L_{H\alpha}$ is the $H\alpha$ luminosity of a SFC and  $L_{H\alpha}(On)$ is the $H\alpha$ luminosity of single $On$ star, where n is the class (e.g. n=3, O3 type star). P(On) is the randomly generated probability for an On-type star, such that $\sum_{n=3}^{9}P(On)=1$. For the second case, we generated the P(On) only for the combination of  O7, O8, and O9 stars. And for the first case, the P(O8)=1. The number of B0 stars in the SFC derived from the FUV luminosity is,

\begin{equation} \label{eq4}
N(B0)= \frac{L_{FUV} - \sum_{n=3}^{9}(N(On) \times L_{FUV}(On))}{L_{FUV}(B0)}
\end{equation}

\noindent
where $L_{FUV}$ and $L_{FUV}(On)$ are the FUV luminosity of a SFC and a single On-type star, respectively. The $L_{FUV}(B0)$ is the FUV luminosity of the B0 star, and N(On) is the number of On-type stars. The total number of O stars is $N(O)=\sum_{n=3}^{9}N(On)$. The number of B1 stars (N(B1)) in a SFC will then be

\begin{multline} \label{eq5}
N(B1)= \frac{L_{NUV}-(N(B0)\times L_{NUV}(B0))}{L_{NUV}(B1)} \\
- \frac{\sum_{n=3}^{9}(N(On)\times L_{NUV}(On))}{L_{NUV}(B1)}
\end{multline} 
where $L_{NUV}$ and $L_{NUV}(On)$ are the NUV luminosity of a SFC and single On stars, respectively. Also, $L_{NUV}(B0)$ and $L_{NUV}(B1)$ are the NUV luminosities of the B0 and B1 stars, respectively.

We then took the ratio of N(O) to N(B0) ($\frac{N(O)}{N(B0)}$) and the ratio of N(B0) to N(B1) ($\frac{N(B0)}{N(B1)}$) to obtain the IMF indices $\alpha_{1}$ and $\alpha_{2}$ respectively. We assume that the IMF can be represented as a distribution function in linear mass units as,
\begin{equation} \label{eq6}
\frac{dN}{dm}= A m^{-\alpha}
\end{equation}

\noindent
where $\alpha$ is the power law index of the IMF. The number of stars N in the mass range $m_u$ to $m_l$ is
\begin{equation} \label{eq7}
N =A \int_{m_l}^{m_u} m^{-\alpha} dm
\end{equation}

\noindent
and $\frac{N(O)}{N(B0)}$ is given by, 
\begin{equation} \label{eq8}
\frac{N(O)}{N(B0)} =\frac{m_{u}(O)^{1-\alpha_{1}}-m_{l}(O)^{1-\alpha_{1}}}{m_{u}(B0)^{1-\alpha_{1}}-m_{l}(B0)^{1-\alpha_{1}}}
\end{equation}

\noindent
where $m_{u}(O)$ and $m_{u}(B0)$ are the upper mass limit for O and B0 stars. Similarly, the lower mass limits for O and B0 stars are $m_{l}(O)$, $m_{l}(B0)$. $m_{u}$ and $m_{l}$ for the SFCs that are fully populated with O9 to O3 type stars is 21 $M_\odot$ to 100 $M_\odot$ respectively. And for B0 stars, it is 21 $M_\odot$ and 10 $M_\odot$. When we took O7, O8 and O9 stars, the $m_{u}$ and $m_{l}$ is 38 $M_\odot$ and 21 $M_\odot$. When considering only the O8 star, the limits are 31 $M_\odot$ and 21 $M_\odot$. All these values are adopted from \cite{Sternberg_2003}.

We also found $\frac{N(B0)}{N(B1)}$ along with the IMF index $\alpha_{2}$. 

\begin{equation} \label{eq9}
\frac{N(B0)}{N(B1)} =\frac{m_{u}(B0)^{1-\alpha_{2}}-m_{l}(B0)^{1-\alpha_{2}}}{m_{u}(B1)^{1-\alpha_{2}}-m_{l}(B1)^{1-\alpha_{2}}}
\end{equation}

\noindent
The $m_{u}$ and  $m_{l}$ for B1 stars is 10 $M_\odot$ and 3 $M_\odot$ respectively. Using the IMF index for all the SFCs, we obtained the distribution of $\frac{N(O)}{N(B0)}$ and $\frac{N(B0)}{N(B1)}$.

\begin{table}[ht!]
\caption{FUV and NUV luminosities of massive stars}
\begin{center}
\begin{tabular}{lllllll}
\hline
OV star & $T_{eff}$& $log(g)$& $L_{FUV}$   & $L_{NUV}$ \\
Type& (kK)&$(cms^{-2})$&($ergs^{-1}$)& ($ergs^{-1}$) \\
\hline
O3& 50& 5 &4.6$\times 10^{38}$&3.96$\times 10^{37}$ \\
O4& 47.5& 5 & 3.5 $\times 10^{38}$& 3.1$\times 10^{37}$\\
O5& 45 &4.5&2.75$\times 10^{38}$&2.48$\times 10^{37}$ \\
O6&42.5&4.5&2.18$\times 10^{38}$&2$\times 10^{37}$\\
O7& 40& 4.5 &1.67$\times 10^{38}$&1.65$\times 10^{37}$\\
O8&38.5& 4.5&1.27$\times 10^{38}$& 1.29$\times 10^{37}$\\
O9&35.5&4&9.13$\times 10^{37}$&1.01$\times 10^{37}$\\
B0&33.3&4&7.26$\times 10^{37}$& 8.17$\times 10^{36}$\\
B1&25 &3.9&6.2$\times 10^{36}$&8.02$\times 10^{35}$\\
\hline
\end{tabular}
\end{center}
\label{table1}  
    \begin{minipage}{80 mm}
    \textit{Note:} $T_{eff}$ and  $log(g)$ are the effective temperature and surface gravity parameters given to the \cite{castelli2004newgridsatlas9model} model to get FUV luminosities ($L_{FUV}$) and NUV luminosities ($L_{NUV}$). We took distance 9.84 Mpc, and the star's radius as mentioned in \cite{Sternberg_2003}.
    \end{minipage}%
\end{table}

\section{Results}\label{sec:Results}

\subsection{Distribution of SFCs in NGC 628 in H\texorpdfstring{$\alpha$}{Lg}, FUV and NUV emission.} \label{Distribution of SFCs in NGC 628 in different wavelengths.}

Although NGC 628 has been observed over the whole spectrum, we focus on the bands that trace star formation in this study, i.e., FUV, NUV, and H$\alpha$. We can interpret the star formation distribution and its propagation by studying the emission at these wavelengths as they trace different timescales. Here, these emissions are characterized as SFCs, which we detected separately from each band. But the number of SFCs detected in each band depends on the threshold we have taken in SExtractor. Although taking a 5$\sigma$ threshold for all bands gives us a confident bright source detection, the different sensitivity of each band may affect the number of SFCs detected and sometimes even the area of the SFCs. However, taking SFCs that are detected in all three bands and taking the same area for all SFCs (the largest area of SFCs of the three bands) for IMF estimation makes us confident of our analysis.

The $3\sigma$ flux sensitivity of MUSE H$\alpha$ is around $4 -7 \times 10^{37} erg s^{-1} kpc^{-2}$ \citep{Eric}. With a 5$\sigma$ threshold, we could detect around 560 SFCs in this band. The H$\alpha$ emission is more prominent in the arm and interarm regions. More than 200 SFCs are detected in the H$\alpha$ map but have no associated UV emission. Some of the H$\alpha$ emission could be associated with diffuse ionized gas (DIG), as indicated in previous studies \citep{Kreckel_2016}. 

The 5$\sigma$ photometric limit of the UVIT F148W filter is 22.78. The FUV SFCs detected are the least numerous (around 190 SFCs) compared to the other two bands with a 5$\sigma$ threshold. As mentioned above, one of the reasons could be the difference in sensitivity, which might have made us lose some faint SFCs. We see that more than 96$\%$ of the FUV SFCs are associated with SFCs in the other two bands, where only six SFCs are associated with FUV in the interarm region (out of 100 interarm SFCs). The reasons for the other 4$\%$ of FUV SFCs not having associated emissions in other bands could be due to the way SFCs are detected, such that associated SFCs could have a center that is a bit farther than the FUV SFCs. Besides this, the shock-excited accretion disk around extremely eruptive young stars emits mainly in FUV and might lack in the other two bands\citep{Carvalho_2024}. Additionally, the enhancement in H$\alpha$ emission from the interarm regions indicates that there could be young SFCs embedded in dust, which obscures the FUV emission. The FUV bright SFCs are also associated with the spurs along the spiral arms, which are prominent in the ALMA images.

The NUV image has brighter SFCs and a smooth diffuse emission over the entire field. The galaxy center is especially bright in NUV, which we do not see in the other two bands. More NUV emission indicates that the galaxy center has older and cooler stellar populations and fewer massive O and B stars. The 5$\sigma$ photometric limit of the UVIT N263M filter is 23.36. Of the 650 SFCs with NUV emission, more than 200 SFCs emit only NUV emission. These could be associated with evolved SFCs. The other SFCs are associated with either FUV or H$\alpha$  emission or both. 
 
\subsection{UV and H\texorpdfstring{$\alpha$}{Lg} luminosity correlation} \label{FUV and Halpha luminosity correlation}

Any IMF study involving UV and H$\alpha$ luminosities is susceptible to extinction correction \citep{Boselli_2009}. Hence, choosing a method for the extinction correction is critical. When we compared the correlation of extinction-corrected luminosities using different methods, we found more scatter for luminosities corrected for extinction using the Beta slope method as seen in Figure \ref{fig2}b and \ref{fig2}e. Previous studies such as \cite{Lee_2009, Hao_2011} have shown that extinction-corrected UV and H$\alpha$ luminosities have a strong correlation. Hence, we obtained the least square fit for UV (FUV and NUV) and H$\alpha$ luminosities, which is corrected for extinction using the Balmer decrement method. This correlation is shown in Figure \ref{fig2}c and  \ref{fig2}f for FUV and NUV, respectively. The equations we get by fitting a regression are

\begin{multline} 
\label{eq_fuv}
log(L(FUV))=(0.933 \pm 0.034)log(L({H\alpha}))\\
+(-0.684 \pm 1.409)
\end{multline} 

\begin{multline} 
\label{eq_nuv}
log(L(NUV))=(0.980 \pm 0.028)log(L({H\alpha}))\\
+(-0.551 \pm 1.096)
\end{multline} 

\begin{figure}[ht!]
\includegraphics[width=1\linewidth]{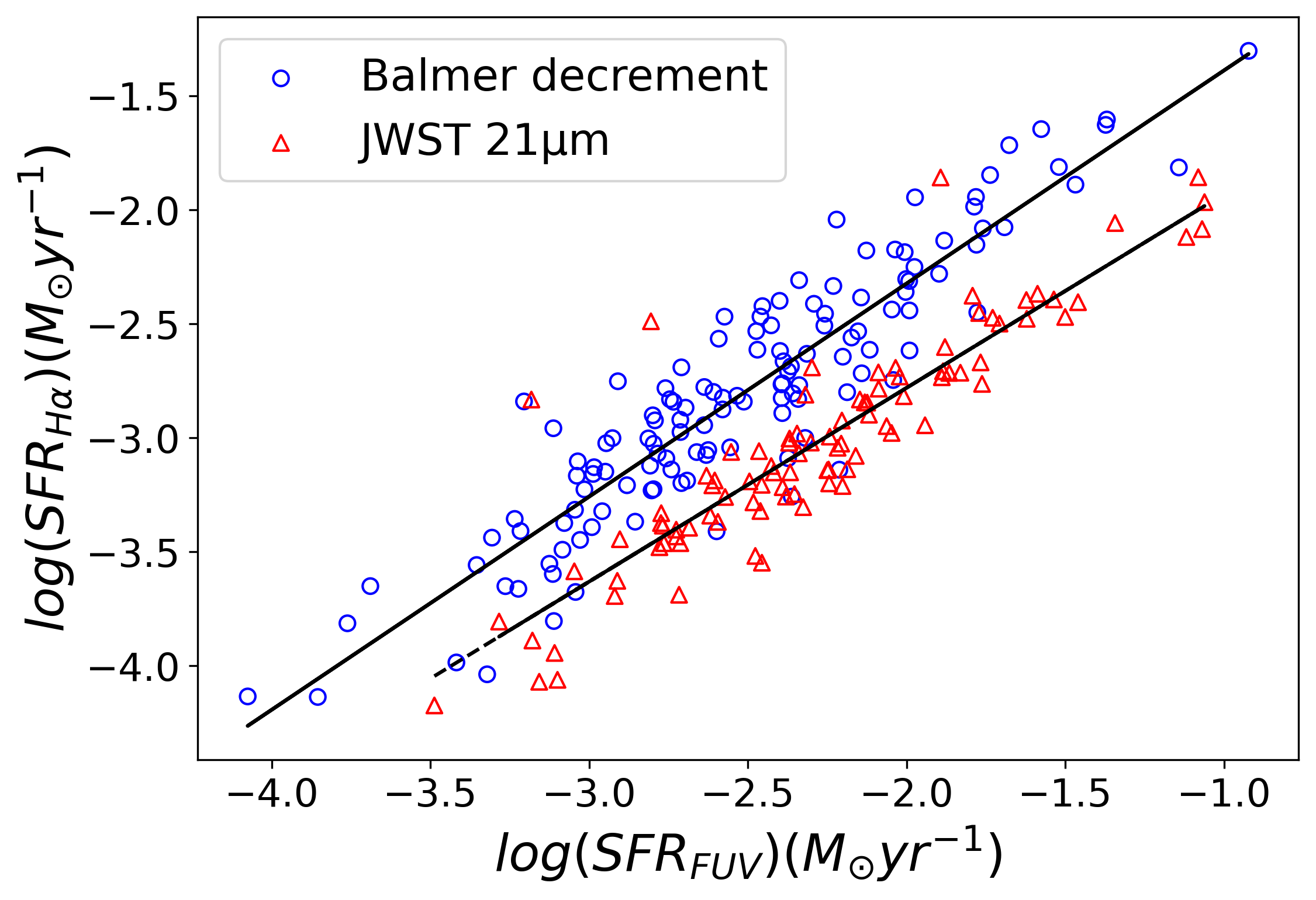}
\caption{Observed H$\alpha$ and FUV SFR correlation for SFCs corrected for extinction with the Balmer decrement and JWST $21 \mu m$ method. The linear fit for the Balmer decrement and JWST $21 \mu m$ method are given as equations \ref{eq_sfr1} and \ref{eq_sfr2}, respectively.}
\label{SFR_correlation}
\end{figure}

Here, equations \ref{eq_fuv} and \ref{eq_nuv} correspond to the correlation of FUV and NUV luminosities with H$\alpha$. All luminosities in equations are in $ergs^{-1}$. We also found a correlation for FUV and H$\alpha$ SFRs for the Balmer decrement and also for JWST 21 $\mu m$ method as shown in Figure \ref{SFR_correlation}. The linear fit for the Balmer decrement is

\begin{multline} 
\label{eq_sfr1}
log(SFR_{FUV})=(0.934 \pm 0.035)log(SFR_{H\alpha})\\
+(-0.455 \pm 0.09)
\end{multline} 
\noindent

and the linear fit for the JWST 21 $\mu m$ method is, 
\begin{multline} 
\label{eq_sfr2}
log(SFR_{FUV})=(0.849 \pm 0.043)log(SFR_{H\alpha})\\
+(-1.081 \pm 0.101)
\end{multline} 
\noindent

From Figure \ref{SFR_correlation} and equations \ref{eq_sfr1} and \ref{eq_sfr2}, we can see that the H$\alpha$ SFR estimated for the Balmer decrement is higher than the JWST method. This again indicates that the different extinction methods correct for extinction at different depths.

 We also see that the SFCs in the interarm region do not have higher luminosities than those in the arms and spurs (See Figure \ref{fig2}c and d). This could be because interarm SFCs have smaller areas, as seen in Figure \ref{fig3}. The luminosity of SFCs in the arm and spurs spans a wide range of values. 
 
 The estimated extinction corrected H$\alpha$ luminosity of SFCs is between $10^{37} \leq L_{H\alpha} \leq 10^{40}$, which is ten times greater than the values in \cite{Kreckel_2016}. This increase in luminosity could be because we have convolved the H$\alpha$ image to a lower resolution, which can cause some clusters to blend into larger ones and potentially increase the contribution of DIGs.

\subsection{Determing the IMF of the SFCs} \label{IMF of the SFCs}

\begin{figure*}[ht!]
\centering
\includegraphics[scale=0.5]{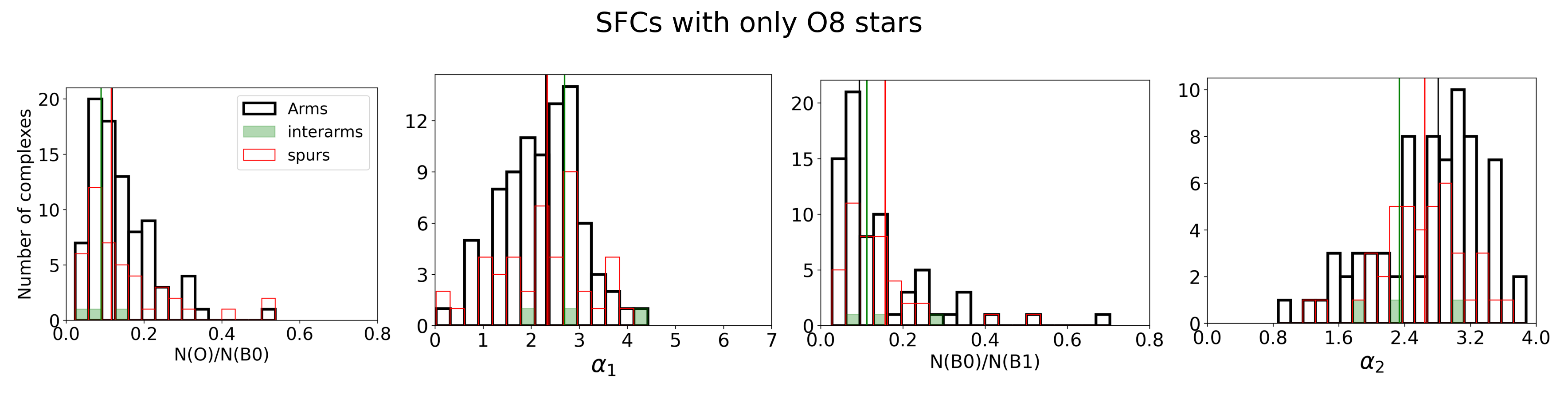}
\includegraphics[scale=0.5]{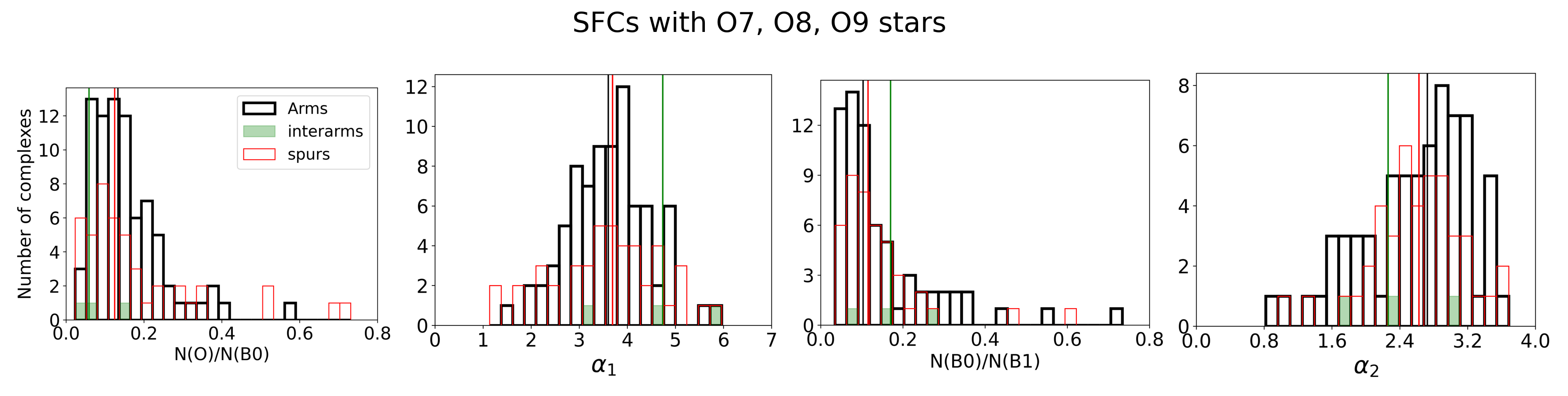}
\includegraphics[scale=0.5]{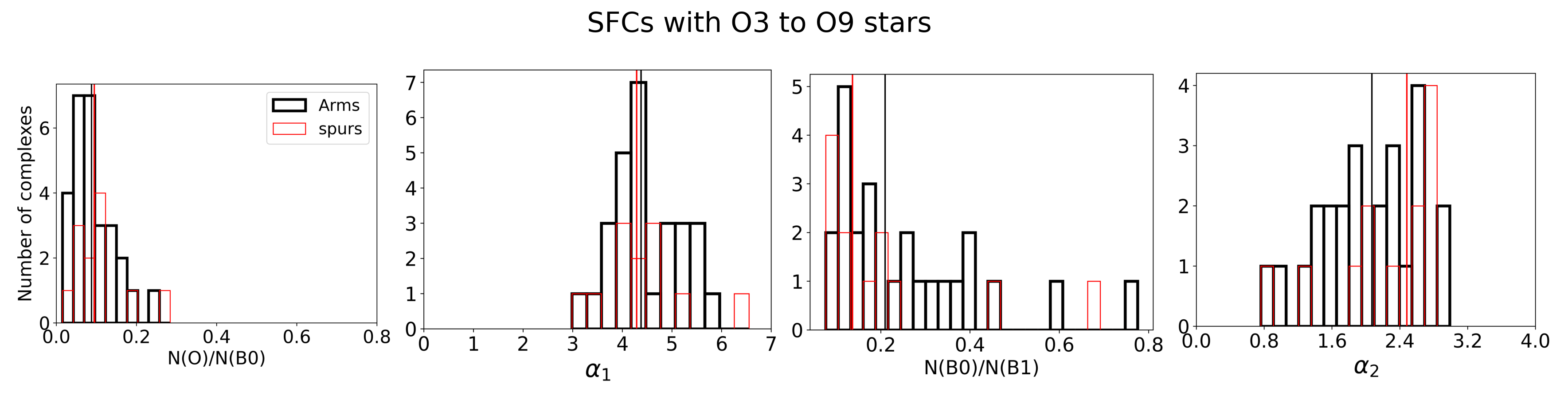}

\caption{The distribution of $\frac{N(O)}{N(B0)}$, $\alpha_{1}$, $\frac{N(B0)}{N(B1)}$ and $\alpha_{2}$. The vertical lines represent the median value of the distribution, which is mentioned in Table \ref{table2}.}
\label{fig4}
\end{figure*}

We estimate $\frac{N(O)}{N(B0)}$ and $\frac{N(B0)}{N(B1)}$ and their respective IMF indices $\alpha_{1}$ and $\alpha_{2}$ in Section \ref{Counts of High mass stars and their ratios}. The distribution of the stellar ratios and IMF indices for the three choices of O-type stars is shown in Figure \ref{fig4}, and the median ratios and IMF index for all the cases are given in Table \ref{table2}. We used the bootstrapping method to determine the standard error associated with the median ratios and median IMF indices.

When we consider that H$\alpha$ arises from only the O8V stars, we find that $\alpha_{1}=2.32$ for the arms and spurs, which is closer to the canonical stellar IMF ($\alpha=2.3$) \citep{kroupa2024initialmassfunctionstars}. But when we populate with more types of O stars, such as O7V, O8V, and O9V, and calculate the median ratio for each SFC, we find that the median $\alpha_{1}$ becomes steeper ($\alpha_{1}=3.66$) than the canonical IMF, and for the arms and spurs, again it is nearly the same. We also noticed that only 2\% of SFCs decreased when populated with three types of O stars (O7V, O8V, O9V), but when we considered the range of O stars (O3V to O9V), we observed a significant decrease in the number of SFCs (69\%). Also, the number of SFCs reduced to 34\% in the arms and 27\% in the spurs, giving the slope a much steeper value of ($\alpha_{1}=4.35$), with no contribution from interarm SFCs. The decrease in the number of SFCs is because not all SFCs have luminosities large enough to accommodate the different types of massive stars. 

However, in contrast to $\alpha_{1}$,  we find that $\frac{N(B0)}{N(B1)}$ and $\alpha_{2}$ do not vary much with different populations of O-type stars. The variation in $\alpha_{2}$ is significantly less for the first two cases. But, we see that $\alpha_{2}$ is steeper for the arms compared to the interarm SFCs, which is contradictory to what is expected. However, this result is statistically insignificant as we have only three SFCs for the first two cases, and the error is also very high for interarm SFCs. When we populate SFCs with O3V to O9V stars, we find that the SFCs in the arms have IMF slopes that are flatter than the spurs. The arms show a more significant variation in $\alpha_{2}$ for the third case compared to the first two cases (which can be up to 0.74). But variation in $\alpha_{2}$ is lesser for spurs (up to 0.16). Nonetheless, the reduced variation and smaller uncertainties associated with it make $\alpha_{2}$ a more reliable indicator of the IMF index. We have discussed more on this topic in Section \ref{sec:discussion}.

\begin{table*}[ht!]
\caption{IMF parameters in different regions}
\begin{center}
\begin{tabular}{llllllll}
\hline
O star & Region   & No of SFCs &  Median $\frac{N(O)}{N(B0)}$  & Median $\alpha_{1}$ & No of SFCs &  Median $\frac{N(B0)}{N(B1)}$ & Median $(\alpha_{2})$ \\
Type && &&  &  & &\\
\hline
&Arm      & 84         & 0.12 $\pm$ 0.01 & 2.30 $\pm$ 0.12          & 71         & 0.09 $\pm$ 0.01 & 2.81 $\pm$ 0.09          \\
O8&Spurs    & 44         & 0.12 $\pm$ 0.01 & 2.33 $\pm$ 0.18          & 42         & 0.11 $\pm$ 0.01 & 2.64 $\pm$ 0.10          \\
&Interarm & 3          & 0.09 $\pm$ 0.04 & 2.69 $\pm$ 0.88          & 3          & 0.16 $\pm$ 0.07 & 2.34 $\pm$ 0.44          \\
\hline
(Weighted Mean Value) &       &            & 0.12 $\pm$ 0.00  & 2.32 $\pm$ 0.06       &            & 0.10 $\pm$ 0.01 & 2.74 $\pm$ 0.10\\
  \hline
&Arm      & 80         & 0.13 $\pm$ 0.01 & 3.60 $\pm$ 0.12               & 67        & 0.10 $\pm$ 0.01 & 2.72 $\pm$ 0.10          \\
O7, O8, O9& Spurs    & 44        & 0.12 $\pm$ 0.02 & 3.69 $\pm$ 0.19 & 42         & 0.12 $\pm$ 0.01 & 2.62 $\pm$ 0.10         \\
&Interarm & 3 & 0.06 $\pm$ 0.05 & 4.73 $\pm$ 0.94                 & 3          & 0.17 $\pm$ 0.08 & 2.26 $\pm$ 0.44          \\
\hline
(Weighted Mean Value) &       &            & 0.12 $\pm$ 0.01 & 3.66 $\pm$ 0.18          &            & 0.11 $\pm$ 0.01 & 2.67 $\pm$ 0.08\\
\hline

\hline
&Arm      & 28         & 0.09 $\pm$ 0.01 & 4.38 $\pm$ 0.19         &   24       & 0.21 $\pm$ 0.04 & 2.07 $\pm$ 0.20 \\
O3 to O9& Spurs    & 12        & 0.09 $\pm$ 0.02 & 4.29 $\pm$ 0.24         & 12         & 0.14 $\pm$ 0.05 & 2.48 $\pm$ 0.27          \\
&Interarm & 0          &-  &  -    & 0         & &    -     \\
\hline

 (Weighted Mean Value) &       &            & 0.09 $\pm$ 0.00 & 4.35 $\pm$ 0.05           &            & 0.19 $\pm$ 0.04 & 2.21 $\pm$ 0.21\\
 \hline


\hline
\end{tabular}
\end{center}
\label{table2}
\hspace{1cm} \begin{minipage}{180 mm}
    \textit{Note:} The weighted mean $\alpha_{1}$ and $\alpha_{2}$ for all three cases combined is 3.16 $\pm$ 0.62 and 2.64 $\pm$ 0.14 respectively.
    \end{minipage}%
\end{table*}

\subsection{IMF trends with the properties of the SFCs} \label{IMF trends with the properties of SFCs}

\begin{figure*}[ht!]

\includegraphics[scale=0.45]{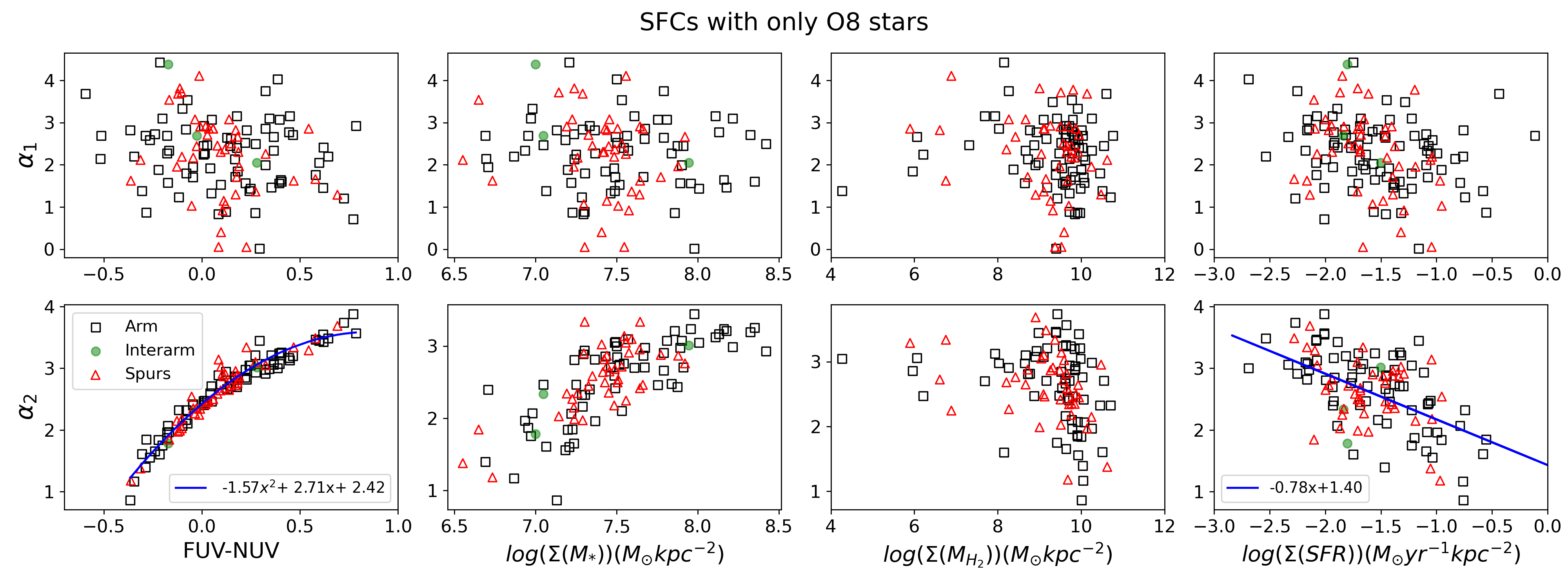}
\includegraphics[scale=0.45]{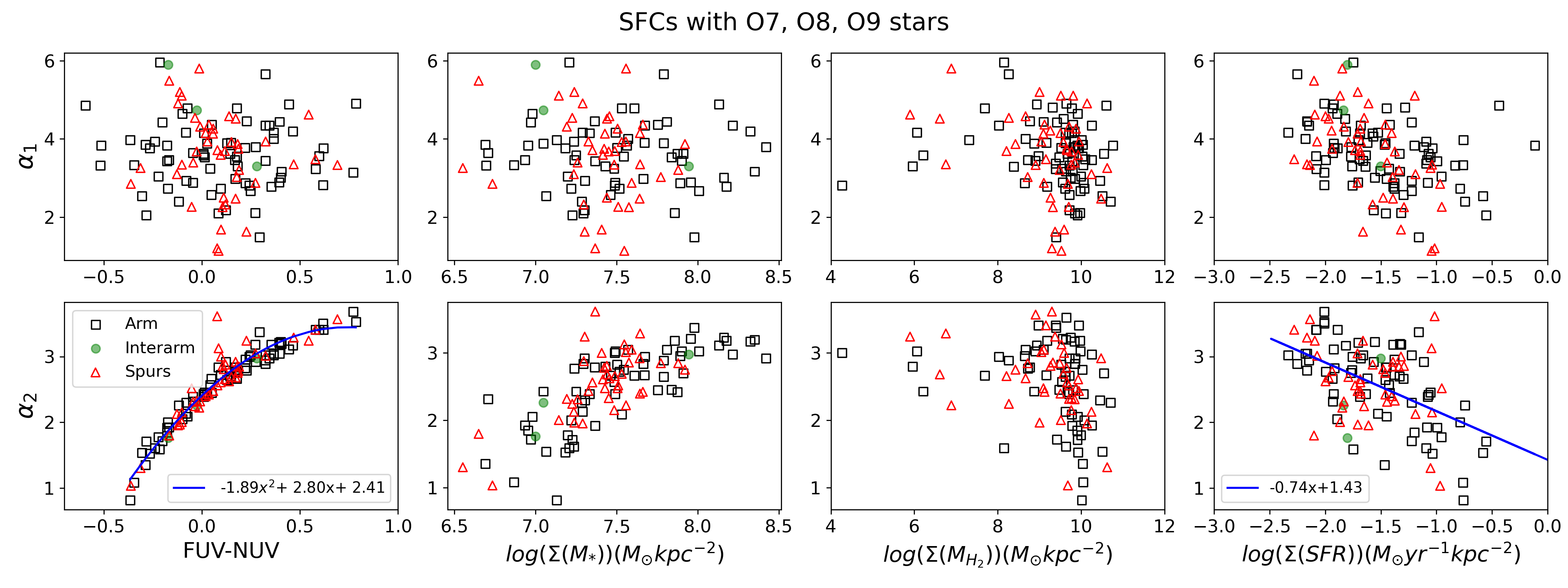}
\includegraphics[scale=0.45]{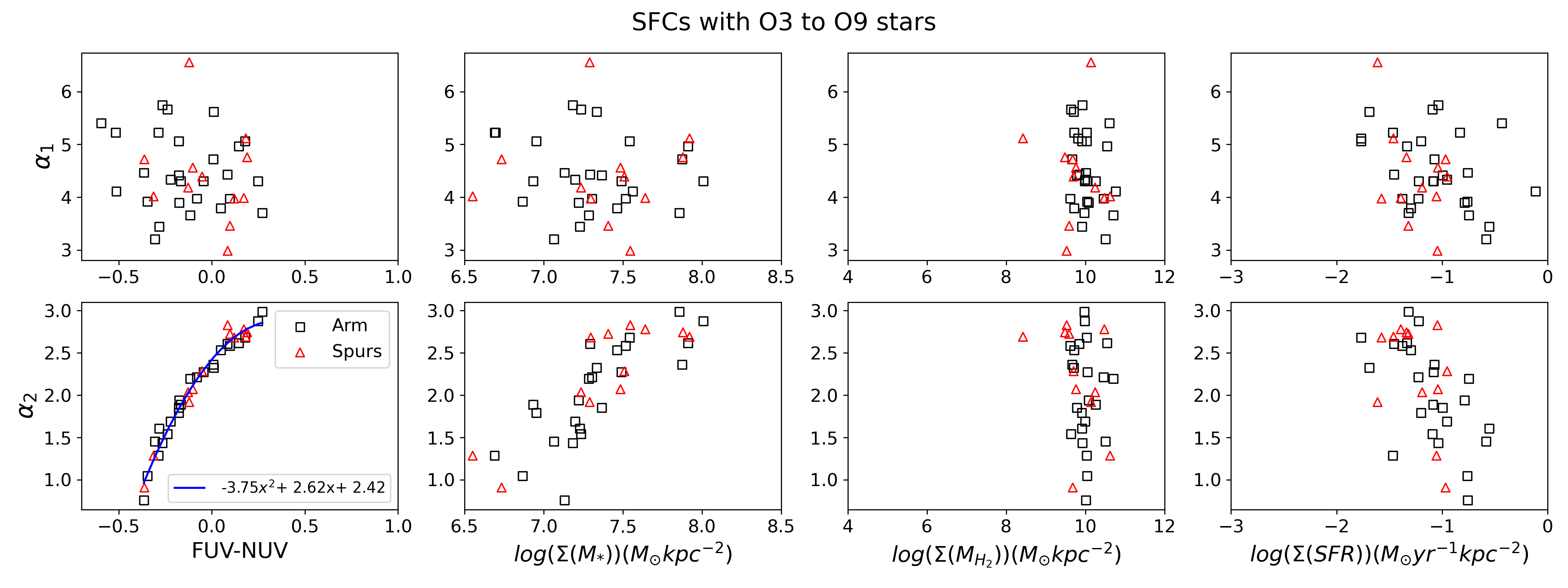}

\caption{Variation of IMF index ($\alpha_{1}$ and $\alpha_{2}$) with the different properties of IMF like FUV and NUV color, stellar mass density, molecular mass density, and star formation rate density (from left).}
\label{fig6}
\end{figure*}

We plotted the variation of our calculated IMF indices ($\alpha_{1}$ and $\alpha_{2}$) with the properties of the SFCs as shown in Figure \ref{fig6}. For the first case (SFCs with only O8 stars), the correlation of with $\alpha_{1}$ with SFC properties such as FUV-NUV color, $log_{10}(\Sigma(M_{*}))$, log($log_{10}(\Sigma(M_{H_2}))$, and $log_{10}(\Sigma(SFR))$ are given by the Spearman correlation coefficient of -0.2, -0.1, -0.1 and -0.3 respectively, which are weak negative correlations. Only with $log_{10}(\Sigma(SFR))$, does $\alpha_{1}$ show a slightly better correlation. We observed that for the second case (SFCs with O9, O8, and O7 stars), the $log_{10}(\Sigma(SFR))$ correlation with $\alpha_{1}$ became stronger with the correlation coefficient -0.5, which indicates that as the SFR increases, the IMF becomes flatter, i.e., the number of massive stars increases. The other three properties show similar correlations with $\alpha_{1}$ for the second case as for the first one.

In contrast, $\alpha_{2}$ shows strong correlations with the properties of the SFCs,  the most prominent being the correlation with FUV-NUV color. It has a high Spearman correlation coefficient value of nearly +1 for all the cases. We fitted a quadratic equation to it as shown below (equation \ref{eq10}), and the fitting parameters are given in Table \ref{tab_3} for the different cases.

\begin{equation}    
 \label{eq10}
\alpha_{2}=a(FUV-NUV)^{2}+b(FUV-NUV)+c
\end{equation}

\begin{table*}[ht!]
\begin{center}
\caption{Fitting parameters for $\alpha_{2}$ and FUV-NUV plot (equation \ref{eq10}) and $\alpha_{2}$ and $log_{10}(\Sigma(SFR))$ plot (equation \ref{eq11}) }
\label{tab_3} 
\begin{tabular}{lllllll}
\hline
O star & a& b& c & d & e\\
Type&&&&($M_{\odot}^{-1}yrkpc^{2}$)& \\
\hline
O8& -1.57 $\pm$ 0.15 & 2.71 $\pm$ 0.07& 2.42 $\pm$ 0.02& -0.78 $\pm$ 0.11& 1.40 $\pm$ 0.18\\
O7, O8, O9& -1.89 $\pm$ 0.18 &2.80 $\pm$ 0.08& 2.41 $\pm$ 0.02&-0.74 $\pm$ 0.12&1.43 $\pm$ 0.19\\
O3 to O9 & -3.75 $\pm$ 0.52 &2.62 $\pm$ 0.11& 2.42 $\pm$ 0.02&-&-\\

\hline
\end{tabular}
\end{center}

\hspace{4cm}    \begin{minipage}{120 mm}
    \textit{Note:} a, b and c are fitting parameters for $\alpha_{2}$ and FUV-NUV plot. Whereas, d and e are fitting parameters for $\alpha_{2}$ and $log_{10}(\Sigma(SFR))$ plot. The first row or only O8 stars gives the best fit for the $\alpha_{2}$ and (FUV-NUV) relation.
    \end{minipage}%
    
\end{table*}

\noindent
The correlation coefficients of $\Sigma(SFR)$ with $\alpha_{2}$ are found to be -0.5 and -0.4 for the first two cases. We also fitted a linear regression for the $\alpha_{2}$ and $\Sigma(SFR)$ correlations. The equation is given by,

\begin{equation} \label{eq11}
\alpha_{2}=d \times log_{10}(\Sigma(SFR))+e
\end{equation}

\noindent
where the fitting parameters d and e are given in Table \ref{tab_3}.

As the canonical IMF is considered a universal probability density distribution function \citep{2018A&A...620A..39J}, we found the corresponding parameters of the SFCs associated with this IMF index so that we can compare it with our estimated IMF. We determined the color and $log_{10}(\Sigma(SFR))$ for the canonical stellar IMF value ($\alpha=2.3$) from equations \ref{eq10} and \ref{eq11} along with Table \ref{tab_3} for all three cases. The color, $FUV-NUV \approx -0.04$ and $log_{10}(\Sigma(SFR)/M_{\odot}yr^{-1}kpc^{-2}) \approx -1.16$ was estimated for the canonical IMF value. The values are similar for all three cases. Since our median IMF indices are steeper than the canonical value for arms and spurs, we obtain $FUV-NUV > -0.04$ and $log_{10}(\Sigma(SFR)/M_{\odot}yr^{-1}kpc^{-2}) < -1.16$ for our values.

Apart from FUV-NUV and $log_{10}(\Sigma(SFR))$, the $log_{10}(\Sigma(M_{H_2}))$ also has a weak negative correlation with $\alpha_2$, with a correlation coefficient of -0.4. However, this is stronger than the correlation with $\alpha_1$. Hence, as $\Sigma(M_{H_2})$ increases $\alpha$ decreases i.e. the IMF becomes flatter. This indicates that the larger the molecular surface mass density in the galaxy disk, the larger are the fraction of high-mass stars. We also see that this correlation generally holds for dense molecular clouds of mass density $\geq 10^{8}M_{\odot}kpc^{-2}$. For less dense clouds, $\alpha$ nearly becomes constant. In Figure \ref{fig6}, in $\alpha_{2}$ and $\Sigma(M_{H_2})$ correlation, there is a fairly sharp cutoff at $log_{10}(\Sigma(H_{2})/M_{\odot}kpc^{-2})=9.5$ for third case. But when it comes to $\Sigma(M_{*})$, we observed the increase in $\alpha_{2}$ with increasing $\Sigma(M_{*})$, which means that the IMF becomes steeper as stellar mass density increases, which is contrary to $\Sigma(M_{H_2})$. This may indicate that the more massive SFCs (i.e., larger $\Sigma(M_{*})$) have evolved stellar populations, and the massive stars have evolved away. We discuss the significance of these trends in Section \ref{sec:discussion}.

\section{Discussion}\label{sec:discussion}

Our study assumes that the H${\alpha }$ emission mainly arises from O-type stars. But, since we have no information on what kind of O stars populate each SFC, we populated the SFCs with the following combinations. (i) O8V star (whose luminosity corresponds to the median luminosity of the SFCs in NGC 628; (ii) a combination of O7V, O8V, and O9V stars and (iii) a combination of OV stars with a mass limit of 100 $M_{\odot}$ (O3V to O9V stars). Then we generated different possible stellar ratios $\frac{N(O)}{N(B0)}$ and $\frac{N(B0)}{N(B1)}$ for each SFC and noted the median values for each SFC. Hence, there is a range of possibilities for the high mass IMF, which depends on how we populate O stars in the SFCs. Apart from O stars, there can also be some contribution from very massive B stars in $H\alpha$, and we are aware of the degeneracy present in considering the type of the population. But for this study, we will go with the above assumption to see if we can get some estimate of the IMF. Even if we consider contribution from B stars in H${\alpha }$ emission, we might see a variation in the result for option (iii), where we consider populating with all types of O stars.

When we populated the SFCs with different types of O stars, we ensured that all types were present in each SFC. But in reality, this may not be the case. But it does help us understand how steep the IMF can go. Studies show that massive O-type stars of mass more than 100 $M_{\odot}$ are detected in nearby galaxies \citep{Kalari_2022}. \cite{koda.etal.2012} have shown that statistically, for the cluster to have a star as massive as 100 $M_{\odot}$, it should have a cluster of mass at least $10^{5}$ $M_{\odot}$. 

But stochastically, massive stars can form in less massive clusters; alternatively, massive stars might not form in massive clusters.
With high stellar density, the inner disk may host clusters with one or more massive stars. Also, it must be noted that although we see in Figure \ref{fig3} that the SFC stellar masses are above $10^{5}$ $M_{\odot}$, the SFCs may be hosting multiple clusters that cannot be individually resolved. However, low-mass O stars are more likely to form rather than high-mass stars. We also find that the number of SFCs decreases when we try to populate with higher masses of O stars. So, we conclude that the existence of SFCs with very massive O-type stars in NGC 628 is less probable than the existence of low mass O-type stars like O9 to O7 stars.

When considering fully populated SFCs (option (iii)), the IMF becomes top-heavy (i.e., $\alpha>2.3$, so a steeper IMF) compared to only for O8V stars (option (i)), which is nearly canonical. Hence, the IMF index is steeper than the canonical value at the high mass end of stellar populations (O stars and massive B stars). But we know that the more massive the stars are, the shorter is their lifetime. Hence, many massive stars must have already passed their main sequence, which could be why the present-day IMF that we observe in NGC 628 appears to be a top-light IMF.  

We have derived two IMF indices $\alpha_1$ and $\alpha_2$. Since $\alpha_1$ is derived from O-type stars that have short lifetimes ($<10^{7}$yrs), it represents the present-day IMF index. On the other hand,  $\alpha_2$ is more dependent on the B-type stars, which have longer lifetimes ($\leq10^{8}$yrs) and hence represent the general high-mass IMF index. However, some emissions from NUV might come from the evolved stellar populations. But as the contribution from main sequence stars will be higher, and since we are taking the SFCs that have emission in other bands too, we can say that the SFCs we have considered have massive stars in them. The $\alpha_1$ values in Table \ref{table2} show a significant variation for different populations of O stars (cases (i), (ii), and (iii)). In contrast, the ratio of high mass to low mass B stars, i.e., N(B0)/N(B1), gives a consistent value for all the scenarios. This shows that $\frac{N(B0)}{N(B1)}$ does not depend strongly on what kind of O stars are present in the SFCs. This suggests that the average of $\alpha_2$ could be a better indicator of the high mass stellar IMF than the present-day IMF index $\alpha_1$. The mean value for the IMF index we obtain is $\alpha_{2}= 2.64 \pm 0.14$, which is a steeper or top-light IMF compared to the canonical IMF.

Various studies have obtained a top light IMF compared to the canonical IMF for the massive stars \citep{2002Sci...295...82K}. The \cite{1979ApJS...41..513M} and \cite{1983ApJ...272...54K} predicted $\alpha=2.5$ for masses above 1$M_{\odot}$. \cite{1986IAUS..116..451S} predicted $\alpha$ is between 2.3 and 3.3 for the high mass end. Apart from these, \cite{2015ApJ...806..198W} found $\alpha=2.45$ for 85 clusters in  M31 and  \cite{Wainer_2024} found $\alpha=2.5$ for the 34 clusters in  M33. These studies used the resolved population in clusters to find the high mass IMF. Most of these studies have the low mass cutoff for the high mass IMF as 1$M_{\odot}$. In our study, there is an overlap in the mass for both  $\frac{N(O)}{N(B0)}$ and $\frac{N(B0)}{N(B1)}$ ratios, and we see that both $\alpha_{1}$ and $\alpha_{2}$ have an average value within the range predicted in \cite{1986IAUS..116..451S}.

Studying the stellar IMF in external galaxies is crucial as it allows us to investigate the universality of the IMF within the galaxy as well. \cite{10.1111/j.1745-3933.2006.00258.x} showed that the IMF in the star-forming regions of the Milky Way bulge differs from the IMF in the disk. Multiple studies utilize the complete field of the galaxy and try to understand if there is any IMF variation in the XUV disk of the galaxy compared to the inner disk \citep{koda.etal.2012, 10.1093/mnras/stz3151}. This is because the IMF can depend on galaxy properties such as stellar densities, gas densities, and metallicity. 

The MUSE observations of NGC 628 cover only the inner region. Hence, we try to understand the properties of SFCs in the arm, interarm, and spurs, which are expected to have different environments and star-forming conditions. Unfortunately, only three interarm SFCs have FUV, NUV, and the $H\alpha$ emission, which is necessary to obtain the IMF index. Hence, the IMF index values we estimated for the interarm region in this study are statistically insignificant. However, as discussed below, our results agree with those of the previous studies. 

The SFCs in the interarm regions have smaller areas and mass densities compared to the arms and spurs. Hence, it is expected that the IMF in the interarms will not have enough cluster mass to have more massive stars, making it top-light IMF  \citep{kroupa2024initialmassfunctionstars}. However, several studies, including \cite{Kreckel_2016}, have shown that there is little difference in star formation properties between arm and interarm SFCs. Our studies show that the IMF value for the interarm region is steeper than the arm and spur regions for $\frac{N(O)}{N(B0)}$, and for $\frac{N(B0)}{N(B1)}$, it is flatter. However, fewer SFCs in the interarm regions make the error bars prominent. \cite{1986IAUS..116..451S} showed that the variation in the IMF with region to region is smaller than $\pm 0.5$. Here, the variation of the IMF index $\ alpha_1$ in the interarm compared to the arm and spurs is less than 0.5 for case one, but it becomes more significant for case two. Apart from the three SFCs in the interarm for which we found the high mass stars ratio, there are more SFCs with only H$\alpha$ and only NUV emission and no counterparts in the two wavelengths. This shows that there could be stars in the very early stage surrounded by gas and dust, causing extinction in lower wavelengths, and the presence of only NUV in some SFCs indicates that there are even older populations in this region.

There is no statistically significant difference between $\alpha_{1}$ in the spurs and the arms for any three cases. The present-day IMF $\alpha_{1}$ of the SFCs in the spurs is slightly steeper than the IMF in the arms. The variation is between 0.03 to 0.09. It becomes flatter for $\alpha_{2}$, making the arms top light than the spurs and interarms. Spurs have nearly the same IMF index for the third case as the first two, but arms become much flatter, making it top heavy, contrary to what we see for $\alpha_{1}$. So $\alpha_{1}$ might be the right indicator to check for the universality and compare the nature of high mass IMF across the region, as how steep or flat $\alpha_{2}$ can get, might be dependent on how we are populating the O stars, though variation in values for different cases is lesser in $\alpha_{2}$.

The variation of high mass IMF indices with the properties of SFCs like SFR, metallicity, and mass of the clusters or galaxies helps us understand the massive star formation in different environments. The $(\Sigma(SFR))$ variation with the IMF is commonly discussed \citep{2011MNRAS.415.1647G, 2018A&A...620A..39J, kroupa2024initialmassfunctionstars}. Studies have shown that the higher the galaxy or cluster star formation rate (SFR), the more likely it will form high-mass stars, as they will have enough mass to produce them \citep{larson2006understandingstellarinitialmass, 2011MNRAS.415.1647G}. We see that our SFCs also follow a similar trend. Even $\alpha_{1}$, which does not show much trend with any other properties, shows the trend with $log_{10}(\Sigma(SFR))$. We fitted a linear regression to the  $\alpha_{2}$ and $log_{10}(\Sigma(SFR))$, and found the relation as in equation \ref{eq11} with the parameters mentioned in Table \ref{tab_3}. \cite{2011MNRAS.415.1647G} also fitted a linear fit for $\alpha$ and $log_{10}(\Sigma(SFR))$ for galaxies, which is

\begin{equation} \label{eq:gunawardhana}
\alpha=-0.3 \times log_{10}(\Sigma(SFR))+1.7
\end{equation}

\noindent{where $\alpha$ is the IMF index as mentioned in the Introduction \ref{sec:intro}. We find that this relation gives $log_{10}(\Sigma(SFR))$ for the $\alpha$ corresponding to the canonical IMF is equal to -2$M_{\odot}yr^{-1}kpc^{-2}$, which is lower than what we obtain for our SFCs for high mass $\alpha$. This implies that the $log_{10}(\Sigma(SFR))$ for the SFCs to produce the canonical IMF is greater than what we see for galaxies. The $log_{10}(\Sigma(SFR)/M_{\odot}yr^{-1}kpc^{-2})$ of the inner regions of the Milky Way varies between 0 and -3 \citep{Elia_2022}. The $log_{10}(\Sigma(SFR)/M_{\odot}yr^{-1}kpc^{-2})$ corresponding to canonical IMF found using relation from \cite{2011MNRAS.415.1647G} and even our values falls within this range. However, in this study, we estimate for SFCs and not for the whole galaxy.}

We find a strong correlation between $\alpha_{2}$ and the extinction-corrected UV color, FUV-NUV. We see in our distribution in Figure \ref{fig3} that the SFCs have a wide range of color from -0.6 to 0.8. The median color for the SFC distribution in the arms is 0.14, and 0.08 in spurs. \cite{koda.etal.2012} discussed that if a cluster has $FUV-NUV < 0.2-0.3$, ie, is blue in color, it must contain O and/or B stars. However, it is difficult to differentiate clusters with and without O stars just by looking at the UV colors. Even SFCs with $FUV-NUV>0.3$ might have O and B stars. Hence, we took blue SFCs with $FUV-NUV \leq 0.3$. We found the median O to B0 and B0 to B1 stellar ratios, along with median $\alpha_{1}$ and $\alpha_{2}$ values as seen in Table \ref{table_discussion}. We found that it significantly affects the arm SFC numbers. For the first two cases, the number of SFCs is the same. We also found that taking only blue SFCs does not affect $\alpha_{1}$ much. However, $\alpha_{2}$ values became less steep, making $\alpha_{2}$ closer to the canonical value. For the first two cases, the arms and spurs have nearly the same $\alpha_{2}$ values. We also find that spurs become steeper than arms for $\alpha_{2}$. We found that the IMF index and FUV-NUV correlations are strong only for $\alpha_{2}$. One of the reasons for this could be because we obtain the number of B stars from the FUV and NUV emissions. 

\begin{table*}[ht!]
\caption{IMF parameters in different regions for SFCs with FUV-NUV $\leq$ 0.3.}
\begin{center}
\begin{tabular}{llllllll}
\hline
O star & Region   & No of SFCs &  Median $\frac{N(O)}{N(B0)}$  & Median $\alpha_{1}$ & No of SFCs &  Median $\frac{N(B0)}{N(B1)}$ & Median $(\alpha_{2})$ \\
Type && &&  &  & &\\
\hline
&Arm      & 51         & 0.12 $\pm$ 0.01 & 2.27 $\pm$ 0.15          &   48       & 0.14 $\pm$ 0.02 & 2.47 $\pm$ 0.10         \\
O8&Spurs    & 37         & 0.11 $\pm$ 0.02 & 2.36 $\pm$ 0.18          & 37         & 0.12 $\pm$ 0.01 & 2.54 $\pm$ 0.09          \\
&Interarm & 3          & 0.09 $\pm$ 0.04 & 2.69 $\pm$ 0.88          & 3          & 0.16 $\pm$ 0.08 & 2.34 $\pm$ 0.44          \\
\hline
(Weighted Mean Value) &       &            & 0.12 $\pm$ 0.01  & 2.32 $\pm$ 0.08       &            & 0.13 $\pm$ 0.01 & 2.49 $\pm$ 0.04\\
  \hline
&Arm      & 51         & 0.14 $\pm$ 0.01 & 3.55 $\pm$ 0.13               & 48        & 0.14 $\pm$ 0.02 & 2.43 $\pm$ 0.11          \\
O7, O8, O9& Spurs    & 37        & 0.12 $\pm$ 0.02 & 3.68 $\pm$ 0.22 & 37         & 0.12 $\pm$ 0.01 & 2.54 $\pm$ 0.09         \\
&Interarm & 3 & 0.06 $\pm$ 0.05 & 4.73 $\pm$ 0.94                 & 3          & 0.17 $\pm$ 0.08 & 2.26 $\pm$ 0.44          \\
\hline
(Weighted Mean Value) &       &            & 0.13 $\pm$ 0.02 & 3.64 $\pm$ 0.20          &            & 0.13 $\pm$ 0.01 & 2.47 $\pm$ 0.06\\
\hline

\hline
&Arm      & 27         & 0.09 $\pm$ 0.01 & 4.33 $\pm$ 0.18         &   24       & 0.21 $\pm$ 0.04 & 2.07 $\pm$ 0.20 \\
O3 to O9& Spurs    & 12        & 0.09 $\pm$ 0.02 & 4.29 $\pm$ 0.24         & 12         & 0.14 $\pm$ 0.05 & 2.48 $\pm$ 0.27          \\
&Interarm & 0          &-  &  -    & 0         & &    -     \\
\hline

 (Weighted Mean Value) &       &            & 0.09 $\pm$ 0.00 & 4.32 $\pm$ 0.05           &            & 0.19 $\pm$ 0.04 & 2.21 $\pm$ 0.21\\
 \hline


\hline
\end{tabular}
\end{center}
\label{table_discussion}
\hspace{1cm} \begin{minipage}{180 mm}
    \textit{Note:} The weighted mean $\alpha_{1}$ and $\alpha_{2}$ for all three cases combined is 3.22 $\pm$ 0.6 and 2.43 $\pm$ 0.06 respectively.
    \end{minipage}%
\end{table*}

The color $FUV-NUV > 0$  implies that the NUV emission is more than the FUV. That is, there are fewer massive stars. NUV also comes from some evolved stars that can contribute and make SFCs redder. This can also be seen in the trend. The IMF gets steeper as the SFCs become redder, again showing the bottom heaviness. And for the cases, it starts getting bottom heavy at $FUV-NUV>-0.04$. Hence, color can also indicate the nature of the IMF.

From our results, we also find that the weak correlation of $\alpha_{2}$ with $\Sigma(M_{H_2})$ is as expected, i.e., as molecular hydrogen mass density increases, the IMF index increases. The more massive and dense the clouds, the higher the probability of forming high-mass stars. The reason for the scatter in the correlation could be that a single conversion factor was considered to obtain the molecular hydrogen mass for all the SFCs. It also seems like the less dense clouds cannot fully populate the O stars. In Figure \ref{fig6}, for case three, molecular clouds barely form massive O stars for $log(\Sigma(M_{H_2})/M_{\odot}kpc^{-2})<9.5 $ and none below $log(\Sigma(M_{H_2})/M_{\odot}kpc^{-2})=8.5$. We also see that most of the data points in all three runs are still clumped between $\alpha_{2}=$2 to 3, as expected from canonical or Salpeter, and some region-to-region variation. The tail to $\alpha_{1}$ only appears at the highest $\Sigma(M_{H_2})$ measured. This shows the importance of molecular cloud surface density in forming massive stars.

The correlation of $\alpha_{2}$ with $\Sigma(M_{*})$ in our plot seems contradictory to the previous studies. As the SFC $\Sigma(M_{*})$  increases, the IMF steepens, implying fewer massive stars are produced. However, we estimated SFC masses from the starburst99 model, and the redder the SFCs, the more massive the complexes are, i.e., they are composed of cooler and older stars. But this does not mean that there were no massive stars before. As said before, the more massive the stars are, the shorter their time in the main sequence will be. Even if these SFCs had very massive stars, they might already be exhausted. Hence, this trend is quite understandable. However, a more accurate estimation of stellar mass might help us understand it better.

We find very few FUV excess young SFCs with low stellar mass that show top-heavy IMF (see $\alpha_{2}$ correlation with FUV-NUV in Figure \ref{fig6}.) There are very few in arms and spurs. Some even show nearly Salpeter IMF. These SFCs have large $\Sigma(SFR)$. This could be an indication of recent starbursts in these SFCs. 

Apart from these factors, the IMF also depends upon the metallicity and temperature of the molecular clouds. As we look at the inner disk of a massive galaxy, we expect solar metallicity to be a reasonable assumption, and we do not expect it to vary much. Although not knowing the exact population is a major drawback, this study gives an idea of the IMF in galaxies where the $H\alpha$ mainly comes only from O stars. This result is specific to the galaxy NGC 628, which is a massive grand design spiral galaxy. In future studies, we will apply the same method to explore the IMF in galaxies in different environments and check its universality. 

\section{Conclusions} \label{sec:Conclusions}
In this project, we used H$\alpha$ from MUSE and FUV and NUV emission from UVIT to determine the IMF of the SFCs in the inner disk of the galaxy NGC 628 by estimating the approximate number of O stars and B stars from each SFC. We also derived some important trends of the IMF index with the properties of the SFCs. The main results of this study are summarised below:

1. There is less FUV emission in the inner disk of NGC 628 compared to $H\alpha$ and NUV emission, indicating strong extinction and/or more dust-embedded star formation, mainly in the interarm regions.

2. We see a strong correlation between the extinction-corrected UV and H$\alpha$ luminosities for the Balmer decrement method and 21 $\mu$m JWST correction compared to the extinction correction done by the Beta slope method.

3. The properties of SFCs in the arm, spurs, and interarms do not vary much and are consistent with previous studies. However, the area of the SFCs in the interarms is smaller compared to the distribution of SFC areas in arms and spurs.

4. The present day IMF index $\alpha_{1}$ corresponding to $\frac{N(O)}{N(B0)}$ stellar ratio becomes steeper ($\alpha_{1}=4.35 \pm 0.05$) as we populate the SFCs with O-type stars upto stellar masses of 100 $M_{\odot}$ (option 3, Table \ref{table2}), which is much steeper than the values found in the previous studies. The mean $\alpha_{1}$ is around $3.16 \pm 0.52$. However, the general high mass IMF index $\alpha_{2}$ corresponding to the $\frac{N(B0)}{N(B1)}$ ratio, has mean slope of $\alpha_{2}=2.64 \pm 0.14$. The $\alpha_{2}$ values are consistent for different O-type stellar populations. However, $\alpha_{1}$ for the case where SFCs were populated with only O8 stars has a mean value closer to the canonical stellar IMF.

5. The present-day IMF index $\alpha_{1}$ from massive stars ($M>10 M_{\odot}$) is steeper than high mass IMF index $\alpha_{2}$, which we estimated from stars of mass, $10 M_{\odot} \geq M \geq 3 M_{\odot}$. This is because the more massive the star is, the shorter is its lifetime in the main sequence.

6. The IMF index in the arms and spurs are similar, showing a variation $<\pm 0.2$. The IMF index $\alpha_{1}$ is steeper for the interarm than the IMF index for the arm and spurs. Whereas $\alpha_{2}$ gets steeper for the arm than spurs and interarm.

7. The extinction-corrected UV color (FUV-NUV) strongly correlates with $\alpha_{2}$. But for $FUV-NUV> -0.04$, the IMF becomes steeper. Hence, the UV color can also indicate the nature of the IMF.

8. The median value of $\alpha_{1}$, when considered SFCs with $FUV-NUV \leq 0.3$ is consistent with the value when considered all the SFCs. But the median value of $\alpha_{2}$ becomes less steeper with median value $\alpha_{2}=2.43 \pm 0.06$.

9. The $log_{10}(\Sigma(SFR))$ vs $\alpha_{2}$ shows a similar trend as previous studies for galaxies. But the value $log_{10}(\Sigma(SFR))$ for galaxies corresponding to the Salpeter IMF is smaller than what we see for SFCs. 

10. In our study, $\alpha_{2}$ shows a cutoff at $M_{H_{2}}\sim10^{9.5} M_{\odot}$. On closer inspection, $\alpha_{2}$ increases with the increase in the density of molecular gas with a weak correlation, suggesting that low-density clouds cannot support high-mass stars, and most of the SFCs clump around the canonical IMF value.

11. The $\alpha_{2}$ and $\Sigma(M_{*})$ trend shows that massive and old SFCs must have gone through episodes of star formation. Hence, some high-mass stars must already be in the main sequence turn-off. 

12. Young SFCs ($FUV-NUV< -0.04$) with less stellar mass densities show the top-heavy IMF. They also show the high  $log_{10}(\Sigma(SFR))$, which suggests the starburst in these SFCs.

\section{acknowledgments}
We thank the referee for the valuable comments. MD acknowledges the support of the Science and Engineering Research Board (SERB) Core Research Grant CRG/2022/004531 and the Department of Science and Technology (DST) grant DST/WIDUSHI-A/PM/2023/25(G) for this research. We thank Dr. Kathryn Kreckel for her guidance on using MUSE data. This publication uses the data from the UVIT, which is part of the AstroSat mission of the Indian Space Research Organisation (ISRO), archived at the Indian Space Science Data Centre (ISSDC). We gratefully thank all the members of various teams for supporting the project from the early stages of design to launch and observations in orbit. 

 This work is based on observations made with the NASA/ESA/CSA JWST. The data were obtained from the Mikulski Archive for Space Telescopes at the Space Telescope Science Institute, which is operated by the Association of Universities for Research in Astronomy, Inc., under NASA contract NAS 5-03127. The observations are associated with JWST program 2107. The specific JWST observation analyzed can be accessed via doi:10.17909/52vt-2g08. Based on observations collected at the European Southern Observatory under ESO programs 094.C-0623 (PI: Kreckel), 095.C-0473, 098.C-0484 (PI: Blanc), and 1100.B-0651 (PHANGS-MUSE; PI: Schinnerer), as well as 094.B-0321 (MAGNUM; PI: Marconi), 099.B-0242, 0100.B-0116, 098.B-0551 (MAD; PI: Carollo), and 097.B-0640 (TIMER; PI: Gadotti). This paper makes use of the following ALMA data: ADS/JAO.ALMA\#2012.1.00650.S. ALMA is a partnership of ESO (representing its member states), NSF (USA), and NINS (Japan), together with NRC (Canada), MOST and ASIAA (Taiwan), and KASI (Republic of Korea), in cooperation with the Republic of Chile. The Joint ALMA Observatory is operated by ESO, AUI/NRAO, and NAOJ. The National Radio Astronomy Observatory is a facility of the National Science Foundation operated under a cooperative agreement by Associated Universities, Inc. 

\vspace{5mm}

\facilities{Astrosat(UVIT), VLT/MUSE ($H\alpha$, $H\beta$), JWST ($21\mu m$), ALMA}

\software{Astropy \citep{2013A&A...558A..33A,2018AJ....156..123A}, 
          Source Extractor \citep{1996A&AS..117..393B, 2016JOSS....1...58B}, Topcat \citep{2005ASPC..347...29T}, IRAF \citep{1993ASPC...52..173T}, Matplotlib \citep{2007CSE.....9...90H}, NumPy \citep{harris2020array}, Photutils \citep{larry_bradley_2022_6825092}
          }

\appendix \label{sec:appendix}

We have tabulated the estimated properties of SFCs as Table \ref{A1}. The entire table for all SFCs is available in machine-readable form.

\begin{splitdeluxetable*}{cccccccccccBccccccccc}


\tabletypesize{\footnotesize}


\tablecaption{Catalog of estimated properties of SFCs}

\tablenum{A1}

\tablehead{\colhead{R.A.(J2000)} & \colhead{Decl.(J2000)} & \colhead{Filters} & \colhead{Region} & \colhead{Area} & \colhead{$A_{H\alpha}$} & \colhead{$\Delta A_{H\alpha}$} & \colhead{$L_{H\alpha}$} & \colhead{$\Delta L_{H\alpha}$} & \colhead{$L_{FUV}$} & \colhead{$\Delta L_{FUV}$} & \colhead{$L_{NUV}$} & \colhead{$\Delta L_{NUV}$} & \colhead{$SFR_{H\alpha}$ } & \colhead{$\Delta SFR_{H\alpha}$} & \colhead{$SFR_{FUV}$} & \colhead{$\Delta SFR_{FUV}$} & \colhead{Color} & \colhead{$log(M_{H_{2}})$} & \colhead{$log(M_{*})$} \\ 
\colhead{(deg)} & \colhead{(deg)} & \colhead{} & \colhead{} & \colhead{($arcsec^{2}$)} & \colhead{(mag)} & \colhead{(mag)} & \colhead{($10^{3}L_{\odot}$)} & \colhead{($10^{3}L_{\odot}$)} & \colhead{($10^{6}L_{\odot}$)} & \colhead{($10^{6}L_{\odot}$)} & \colhead{($10^{6}L_{\odot}$)} & \colhead{($10^{6}L_{\odot}$)} & \colhead{($10^{-4}M_{\odot}yr^{-1}$)} & \colhead{($10^{-4}M_{\odot}yr^{-1}$)} & \colhead{($10^{-3}M_{\odot}yr^{-1}$)} & \colhead{($10^{-3}M_{\odot}yr^{-1}$)} & \colhead{(mag)} & \colhead{($M_{\odot}$)} & \colhead{($M_{\odot}$)} } 

\startdata
24.19352898 & 15.75866545 & All & arm & 13 & 1.07 & 0.01 & 203.35 & 0.6 & 18.9 & 1.65 & 2.05 & 0.13 & 41.22 & 0.12 & 7.19 & 0.02 & -0.52 & 8.16 &  5.14 \\
24.18308751 & 15.79047867 & All & arm & 15.4 & 0.43 & 0.03 & 19.24 & 0.31 & 1.6 & 0.17 & 0.27 & 0.02 & 3.9 & 0.065 & 0.61 & 0.02 & -0.05 &  &  5.23 \\
24.14929603 & 15.79338057 & All & arm & 9.9 & 0.88 & 0.01 & 63.25 & 0.34 & 10.64 & 0.9 & 1.54 & 0.09 & 12.83 & 0.07 & 4.06 & 0.02 & -0.2 & 7.75 &  5.3 \\
24.15068082 & 15.79135657 & All & arm & 18.8 & 0.92 & 0.02 & 35.81 & 0.44 & 16.15 & 1.2 & 2.32 & 0.12 & 7.25 & 0.09 & 6.14 & 0.02 & -0.21 & 6.76 &  5.82 \\
24.16830729 & 15.81789135 & All & arm & 18.7 & 1.28 & 0 & 559.33 & 1.2 & 27.97 & 2.82 & 3.76 & 0.24 & 113.32 & 0.241 & 10.63 & 0.02 & -0.28 & 8.52 &  5.84 \\
24.19055832 & 15.75534162 & All & arm & 29.2 & 0.52 & 0.02 & 37.11 & 0.37 & 4.08 & 0.31 & 0.73 & 0.04 & 7.55 & 0.076 & 1.56 & 0.02 & 0.03 & 7.68 &  5.85 \\
24.15250999 & 15.7895322 & All & arm & 22 & 1.11 & 0.01 & 215.11 & 0.91 & 26.11 & 2.1 & 3.58 & 0.19 & 43.59 & 0.187 & 9.93 & 0.02 & -0.27 & 8.6 & 5.86 \\
24.16370025 & 15.79534197 & All & arm & 70.8 & 0.73 & 0.02 & 258.23 & 2.69 & 33.46 & 1.3 & 4.47 & 0.14 & 52.38 & 0.548 & 12.66 & 0.04 & -0.29 & 9.22 & 5.87 \\
24.17619665 & 15.80583753 & All & arm & 20.9 & 1.01 & 0.01 & 180.08 & 0.5 & 23.63 & 1.68 & 3.32 & 0.16 & 36.51 & 0.103 & 8.99 & 0.02 & -0.24 & 8.3 & 5.89 \\
\enddata

\label{A1}

\tablecomments{Here, R.A.(J2000) and Decl.(J2000) are the right ascension and declination of the SFCs. Filters are wavebands in which SFCs are present. The region represents SFC's location. Area is the area of the SFCs in $arcsec^{2}$. $A_{H\alpha}$ and $\Delta A_{H\alpha}$ is the $H\alpha$ extinction coefficient and error associated. $L_{H\alpha}$, $L_{FUV}$ and $L_{NUV}$ are the extinction corrected $H\alpha$, FUV and NUV luminosities respectively. $\Delta L_{H\alpha}$, $\Delta L_{FUV}$ and $\Delta L_{NUV}$ are the error associated with respective luminosites. $SFR_{H\alpha}$ and $\Delta SFR_{H\alpha}$ are the $H\alpha$ Star formation rate and respective errors. Similarly, $SFR_{FUV}$ and $\Delta SFR_{FUV}$ are the FUV star formation rate and respective errors. Color is a difference in extinction-corrected FUV and NUV magnitudes. $log(M_{H_{2}})$ is molecular hydrogen mass. $log(M_{*})$ is the stellar mass of the SFCs. (The entire table for all SFCs is available in machine-readable form.)}


\end{splitdeluxetable*}

\bibliography{sample631}{}
\bibliographystyle{aasjournal}

\end{document}